\documentstyle[pre,aps,eqsecnum,epsfig,psfig]{revtex}

\def\eps{\varepsilon}
\def\const{{\rm const\,}}
\def\Dm{\widetilde{\cal D}_{\mu}}

\def\k{{\bf k}}

\def\A{{\cal A}}
\def\partt{\mbox{\boldmath $\partial$}}

\begin{document}
\draft

\title{\phantom{tilaaaaaaaaaaaaaaaaaaaaaaaaaaaaaaaaaaaaaaaaaaaaaaaaaaaaaaaaaa}\hfill\raisebox{1ex}{\rm Preprint ESI 1313
(2003)}\\
Turbulence with Pressure: Anomalous Scaling of a Passive Vector
Field}

\author{N.V. Antonov$^{1}$, Michal Hnatich$^{2}$,
Juha Honkonen$^{3}$, Marian Jur\v{c}i\v{s}in$^{2,4}$}

\address{
$^{1}$ Department of Theoretical Physics, St.Petersburg University,
Uljanovskaja 1, St.Petersburg, Petrodvorez, 198504, Russia,
\\
$^{2}$ Institute of Experimental Physics, Slovak Academy of Sciences,
Watsonova 47, 04011 Ko\v{s}ice, Slovakia,
\\
$^{3}$ Theory~Division, Department~of~Physical Sciences, P.O.~Box~64
(Gustaf H\"{a}llstr\"{o}min katu 2), FIN-00014, University~of~Helsinki,
Finland,
\\
$^{4}$ N.N.~Bogoliubov Laboratory of Theoretical Physics,
Joint Institute for Nuclear Research, \\
141980 Dubna, Moscow Region, Russia}

\date{\today}

\maketitle

\begin{abstract}
The field theoretic renormalization group (RG) and the operator
product expansion are applied to the model of a transverse
(divergence-free) vector quantity, passively advected by the
``synthetic'' turbulent flow with a finite (and not small)
correlation time. The vector field is described by the stochastic
advection-diffusion equation with the most general form of the
inertial nonlinearity; it contains as special cases the kinematic
dynamo model, linearized Navier--Stokes (NS) equation, the special
model without the stretching term that possesses additional
symmetries and has a close formal resemblance with the stochastic
NS equation. The statistics of the advecting velocity field is
Gaussian, with the energy spectrum $E(k)\propto k^{1-\eps}$ and
the dispersion law $\omega\propto k^{-2+\eta}$, $k$ being the
momentum (wave number). The inertial-range behavior of the model
is described by seven regimes (or universality classes) that
correspond to nontrivial fixed points of the RG equations and
exhibit anomalous scaling. The corresponding anomalous exponents
are associated with the critical dimensions of tensor composite
operators built solely of the passive vector field, which allows
one to construct a regular perturbation expansion in $\eps$ and
$\eta$; the actual calculation is performed to the first order
(one-loop approximation), including the anisotropic sectors.
Universality of the exponents, their (in)dependence on the
forcing, effects of the large-scale anisotropy, compressibility
and pressure are discussed. In particular, for all the scaling
regimes the exponents obey a hierarchy related to the degree of
anisotropy: the more anisotropic is the contribution of a
composite operator to a correlation function, the faster it decays
in the inertial-range. The relevance of these results for the real
developed turbulence described by the stochastic NS equation is
discussed. {\bf Key words}: Fully developed turbulence, Anomalous
scaling, Passive vector advection, Renormalization group, Operator
product expansion.
\end{abstract}


\section{Introduction} \label{sec:Intro}

It has become a commonplace to complain that theoretical understanding
of turbulence remains the last unsolved problem of classical physics.
Of course, the concept of turbulence refers to a great deal of disparate
physical situations (``almost as varied as in the realm of life,''
Ref. \cite{Legacy}, p.1) and any exhaustive and ultimate ``theory of
turbulence,'' of course, can hardly ever be established. There is, however,
a classical ``list'' of phenomena (or, rather, classes of phenomena) that
represent and illustrate the main features of turbulence: existence and
stability of solutions of hydrodynamics equations, convective turbulence,
(in)stability of laminar flows and origin of turbulence, and so on. Those
topics, which are of great practical and conceptual importance, have
always remained in the focus of attention for theoreticians. One of them is
the fully developed (homogeneous, isotropic, inertial-range) hydrodynamical
turbulence. Detailed description of this concept and the bibliography of
this old but still open subject can be found in the classical monographs
\cite{Legacy,Monin,McComb}.

Turbulent flows that occur in various liquids or gases at very high Reynolds
numbers reveal a number of general aspects (cascades of energy or other
conserved quantities, scaling behavior with apparently universal ``anomalous
exponents,'' intermittency, statistical conservation laws and so on), which
support the hopes that those phenomena can be explained within a
self-contained and internally consistent theory. Recent developments in
this area are presented and summarized in Ref. \cite{Wien}.

The most remarkable features of developed turbulence are encoded in the
single term of intermittency. This concept has no rigorous definition
within the classical probabilistic theory; an excellent introduction can be
found in Ref. \cite{Z} and Chap.~8 of book \cite{Legacy}. Roughly speaking,
intermittency means that statistical properties (for example, correlation
or structure functions of the turbulent velocity field) are dominated by
rare spatiotemporal configurations, in which the regions with strong
turbulent activity have exotic (fractal) geometry and are embedded into
the vast regions with regular (laminar) flow.

In the turbulence, such phenomenon is believed to be related to strong
fluctuations of the energy flux. Therefore, it leads to deviations
from the predictions of the celebrated Kolmogorov--Obukhov (KO)
phenomenological theory \cite{Legacy,Monin,McComb}. Such deviations,
referred to as ``anomalous'' or ``non-dimensional'' scaling, manifest
themselves in singular (arguably power-like) dependence of correlation
or structure functions on the distances and the integral (external)
turbulence scale $L$. The corresponding exponents are certain nontrivial
and nonlinear functions of the order of the correlation function, the
phenomenon referred to as ``multiscaling.''

Within the framework of numerous semi-heuristic models the anomalous
exponents are related to statistical properties of the local dissipation
rate, the fractal (Haussdorf) dimension of structures formed by the
small-scale turbulent eddies, the characteristics of nontrivial structures
(vortex filaments), and so on; see Refs. \cite{Legacy,Monin,McComb}
for a review and further references. The common drawback of such models
is that they are only loosely related to underlying hydrodynamical equations,
involve arbitrary adjusting parameters and, therefore, cannot be considered
to be the basis for construction of a systematic perturbation theory in
certain small (at least formal) expansion parameter; see e.g. the remark in
Ref. \cite{SFF}. Thus serious doubts remain about the universality of
anomalous exponents and the very existence of deviations from the KO theory.

The term ``anomalous scaling'' reminds of the critical scaling in models
of equilibrium phase transitions. In those, the field theoretic methods
were successfully employed to establish the existence of self-similar
(scaling) regimes and to construct regular perturbative calculational schemes
(the famous $\eps$ expansion and its relatives) for the corresponding
exponents, scaling functions, ratios of amplitudes etc; see e.g.
\cite{Zinn,book} and references therein.

Here and below, by ``field theoretic methods'' we mean diagrammatic and
functional techniques, renormalization theory and renormalization group,
composite operators and operator algebras (operator-product or
short-distance expansions), instanton calculus and so on.

Of course, the analogy is far from exact. There is a big difference
between the concepts of critical scaling in equilibrium phase transitions
and anomalous scaling in turbulence. Formally speaking, in both cases
one deals with nontrivial powers of the distance, but in the first case
they are divided by the ultraviolet (UV) scale $\ell$, while in the second
the same role is played by the integral, or infrared (IR) scale $L$. It was
hoped that a close analogy can be achieved if the momentum space for
turbulence be confronted with the coordinate space for critical phenomena.
This idea was expressed in a phenomenological ``dictionary,''  where, in
particular, the viscous length $\ell$ (that is, the UV scale of turbulence)
was confronted with the correlation length (that is, the IR scale of critical
phenomena), while the integral scale $L$ was confronted with the molecular
length; see e.g. Refs. \cite{Vok}. Hence the idea of ``inverse''
renormalization group; see Refs. \cite{IKR} for a recent discussion.

The aforementioed phenomenon of multiscaling  was also often opposed to
critical scaling, because in the latter ``everything is determined by
just two exponents $\eta$ and $\nu$.''

It has usually been stressed that the
intermittency is essentially a strongly nonlinear phenomenon, and that,
therefore, the anomalous scaling in turbulence cannot be treated within
any kind of perturbation theory. Probably for this reason (and because
of a very low quality of some related papers) the field theoretic methods,
for many years, have been ignored or taken with a strong skepticism by
the turbulent community. The sharpened formulation of the state of the
art, given in \cite{Legacy}, is that the results obtained by diagrammatic
methods are either wrong or can be derived by much simpler methods
(pp. 214--215). Another, and in a sense opposite, point of view was
expressed in \cite{Burg}: ``...the reason [that the problem of turbulence
is still not solved] lies in the fact that the necessary field theoretic
tools have appeared only recently.''

Although the theoretical description of the fluid turbulence on the basis of
the stochastic Navier-Stokes (NS) equations remains essentially an open
problem, considerable progress has been achieved in understanding simplified
model systems that share some important properties with the real
problem: shell models \cite{Dyn}, stochastic Burgers equation
\cite{Burgulence} and passive advection by random ``synthetic'' velocity
fields \cite{FGV}. Although the shell models, discrete analogs of the
NS equation, exhibit pronounced anomalous scaling, it has mostly been
studied within numerical simulations. The Burgers equation with random or
deterministic initial conditions has been extensively studied analytically,
and it exhibits strong intermittency and the energy cascade. The model is
interesting in itself and has various applications (e.g., description of
the development of singularities in self-gravitating matter), but its
relevance for the real hydrodynamical turbulence is far from obvious.
In particular, Burgulence is  a ``turbulence without pressure'' \cite{Burg}
and, what is more, without the energy conservation (in more than one
dimensions), while the conservation of energy and the energy exchange
between different velocity components are very important features
of the genuine fluid turbulence.

Probably the most important progress in the subject, achieved in the last
decade of the twentieth century, was related to a simplified model of the
fully developed turbulence, the so-called rapid-change model. The model,
which dates back to classical studies of Batchelor, Obukhov, Kraichnan and
Kazantsev, describes a scalar or vector quantity (e.g. temperature,
concentration of admixture particles or a weak magnetic field), passively
advected by a Gaussian velocity field, decorrelated in time and self-similar
in space (the latter property mimics some features of a real turbulent
velocity ensemble).

There, for the first time the existence of anomalous scaling was established
on the basis of a microscopic model \cite{Kraich1}, and
the corresponding anomalous exponents were derived within controlled
approximations \cite{Falk1,GK} and regular perturbation schemes \cite{RG}.
Detailed review of the recent theoretical research on the passive scalar
problem and more references can be found in \cite{FGV}.

It is important to emphasize here, that the two alternative (or
complementary) analytical approaches to the rapid-change model are both
field theoretic. In the ``zero-mode approach,'' developed in \cite{Falk1,GK}
(see also \cite{FGV}), nontrivial anomalous exponents are related to the zero
modes (unforced solutions) of the closed exact differential equations
satisfied by the equal-time correlation functions. From the field theoretic
viewpoint, this is a realization of the well-known idea of self-consistent
(bootstrap) equations, which involve skeleton diagrams with dressed lines
and dropped bare terms (see e.g. Sec. 4.35 in book \cite{book}). Owing to
special features of the rapid-change models (linearity in the passive field
and time decorrelation of the advecting field) such equations are exactly
given by one-loop approximations, and the resulting equations in the
coordinate space are differential (and not integral or integro-differential
as in the case of a general field theory). In this sense, the model is
``exactly soluble.'' Furthermore, in contrast to the
case of nonzero correlation time, closed equations are obtained for the
{\it equal-time correlations}, which are Galilean invariant and, therefore,
not affected by the so-called ``sweeping effects'' that would obscure the
relevant physical interactions.

In this connection, it should be noted that, due to the time decorrelation,
in the rapid-change model there is no problem in relating Eulerian
and Lagrangian statistics of the velocity field: they are identical.
This allows one to perform very accurate numerical simulations in the
Lagrangian frame; see \cite{VMF}.

From a more physical point of view, zero modes can be interpreted as
statistical conservation laws in the dynamics of particle clusters
\cite{slow}. The concept of statistical conservation laws appears rather
general, being also confirmed by numerical simulations of Refs.
\cite{CV,SCL}, where the passive advection in the two-dimensional
Navier--Stokes (NS) velocity field \cite{CV} and a shell model of a
passive scalar \cite{SCL} were studied. This observation is rather
intriguing because in those models no closed equations for
equal-time quantities can be derived due to the fact that the
advecting velocity has a finite correlation time (for a passive
field advected by a velocity with given statistics, closed
equations can be derived only for different-time correlation
functions, and they involve infinite diagrammatic series).

The second systematic analytical approach to the rapid-change model, proposed
in paper \cite{RG}, is based on the field theoretic renormalization group
(RG) and operator product expansion (OPE).

To avoid possible confusion, it should be explained that in Ref. \cite{RG}
and subsequent papers, the conventional renormalization group (and not the
inverse RG in the spirit of Refs. \cite{Vok,IKR}) was employed, which is
based on the standard  renormalization procedure (elimination of UV
divergences). The solution proceeds in two main stages. In the first stage,
the multiplicative renormalizability of the corresponding field theoretic
model is demonstrated and the differential RG equations for its correlation
functions are obtained. The asymptotic behavior of the latter on their UV
argument $(r/\ell)$ for $r\gg\ell$ and any fixed $(r/L)$ is given by IR
stable fixed points of those equations. It involves some ``scaling functions''
of the IR argument $(r/L)$, whose form is not determined by the RG equations.
In the second stage, their behavior at $r\ll L$ is found from the OPE within
the framework of the general solution of the RG equations. There, the crucial
role is played by the critical dimensions of various composite operators,
which give rise to an infinite family of independent scaling exponents
(and hence to multiscaling).

Of course, these both stages (and thus the phenomenon of multiscaling)
have long been known in the RG theory of critical behavior, where the OPE
is used in the analysis of the small-$(r/L)$ form of the scaling functions;
see e.g. \cite{Zinn,book} and references therein. The distinguishing feature,
specific to models of turbulence, is the
existence of composite operators with {\it negative} critical dimensions.
Such operators are termed ``dangerous,'' because their contributions to the
OPE diverge at $(r/L)\to0$. In the models of critical phenomena, nontrivial
composite operators always have strictly positive dimensions, so that they
only determine corrections (vanishing for $(r/L)\to0$) to the leading terms
(finite for $(r/L)\to0$) in the scaling functions (the leading terms are
related to the simplest operator unity with zero critical dimension).

The OPE and the concept of dangerous operators in the stochastic
hydrodynamics were introduced and investigated
in detail in \cite{JETP}; detailed discussion of the NS case
can be found in the review paper \cite{UFN}, the monograph \cite{turbo},
and Chap. 6 of the book \cite{book}. Later, the idea of negative
dimensions was repeatedly introduced in connection with the
anomalous scaling in turbulence \cite{Eyink}, models with multifractal
behavior \cite{DL} and the phenomena related to the Burgers equation
\cite{Burg,Burg1}.

The RG analysis of Ref. \cite{RG} has shown that dangerous operators are
indeed present in the rapid-change model, and that their dimensions
can be calculated
systematically within a regular perturbation expansions, similar to the
famous $\eps$ expansion of the critical exponents. Owing to the linearity
of the original stochastic equations in the passive field, only finite
number of dangerous operators can contribute to any given structure function,
which allows one to identify the corresponding anomalous exponent with the
critical dimension of an individual composite operator. The actual
calculations were performed to the second \cite{RG} and third \cite{RG100}
orders in $\eps$ (two-loop and three-loop approximations, respectively).
Generalizations to the cases of compressible \cite{Komp,Komp2} and
anisotropic \cite{Uniaxi} velocity ensembles and the vector advected field
\cite{Lanotte2,alpha,amodel,LA,vector} have been obtained.

The two approaches complement each other very well: the zero-mode technique
allows for exact (nonperturbative) solutions for the anomalous
exponents related to
second-order correlation functions \cite{Falk1,V96,Lanotte,AB} (they
are nontrivial for passive vector fields or  anisotropic sectors for scalar
fields), while the RG approach form the basis for systematic perturbative
calculations of the higher-order anomalous exponents
\cite{RG,RG100,Komp,Komp2}.
For the cases of anisotropic velocity ensembles or/and passively advected
vector fields, where the calculations become rather involved, all the
existing results for higher-order correlation functions were derived only
by means of the RG approach and only to the leading order in $\eps$
\cite{Uniaxi,Lanotte2,alpha,amodel,LA,vector}.

Besides the calculational efficiency, an important advantage
of the RG approach is its relative universality: it is not bound
to the aforementioned
``solubility'' of the rapid-change model and can
also be applied to the case of finite correlation time or
non-Gaussian advecting field  \cite{RG3,Juha2,RG4}.

It has been usually stressed that intermittency and anomalous scaling in
turbulence are signatures of highly nonlinear nature of underlying dynamics.
The main lesson that can be learned from the rapid-change model is, probably,
that such phenomena can be encountered in a model, which is linear in the
passive field, and in which the advecting velocity field is Gaussian and
nonintermittent (in
contrast to a more realistic case of the stochastic NS equation). What is
more, the RG and OPE approach shows that intermittency (at least in the
rapid-change model) can be essentially a perturbative phenomenon, in the
sense that it is contained completely already in the ordinary (primitive)
perturbation theory around a free (Gaussian) approximation. The
infinite resummation of the primitive perturbation series, performed by
the RG and OPE, gives rise to improved representations of the correlation
functions, which reveal anomalous scaling behavior. On the other hand,
these representations can be expanded back and reproduce the original
perturbation series, no less and no more.

Existence of exact solutions, regular perturbation schemes and accurate
numerical simulations allows one to discuss, for the example of the
rapid-change model and its relatives, the issues that are interesting within
the general context of fully developed turbulence: universality and
saturation of anomalous exponents, effects of compressibility, anisotropy
and pressure, persistence of the large-scale anisotropy and hierarchy of
anisotropic contributions, convergence properties and nature of the $\eps$
expansions, and so on.

So far, however, aforementioned field theoretic methods have had only
limited success when applied to the real fluid turbulence or, better
to say, to the stochastic NS equation.

The main problem of the self-consistency approach to the stochastic NS
equation is the elimination of the kinematic sweeping effects, which
obscure relevant physical interactions and lead to a spurious strong
dependence of the correlation functions on the integral scale.
This problem was  claimed solved using the so-called internal diagrammatic
technique \cite{Lvov}, but it leads to the violation of the translational
invariance. Probably for this reason no attempts have been made to
explicitly solve the resulting equations at least in the simplest
one-loop approximation.

The standard RG approach to the stochastic NS equation allows one to prove
the independence of the inertial-range correlation functions of the viscous
scale (the second Kolmogorov hypothesis) and calculate a number of
representative constants within a regular $\eps$ expansions in a reasonable
agreement with experiment; see e.g. \cite{book,UFN,turbo} for a review and
Refs. \cite{APS} for the most recent results. The problem of the anomalous
scaling, that is, the dependence of the Galilean invariant correlation
functions on the IR scale $L$, still remains open, probably due to the lack
of an appropriate expansion parameter. Dangerous operators in that model are
absent in the $\eps$ expansions and can appear only at finite values of
$\eps$. This means that they can be reliably identified only if their
dimensions are derived {\it exactly} with the aid of Schwinger equations or
Galilean symmetry. Due to the nonlinear nature of the problem, they enter the
corresponding OPE as infinite families whose spectra of dimensions are not
bounded from below, and in order to find their dependence on the IR scale $L$
one has to sum up all their contributions. The needed summation of the most
singular contributions, related to the powers of the velocity field (their
critical dimensions are known exactly), was performed in \cite{JETP} with
the aid of the so-called infrared perturbation theory for the case of the
different-time pair correlation functions. It has revealed their strong
dependence on $L$, that physically can be explained by the aforementioned
sweeping effects. This demonstrates that, contrary to what is sometimes
claimed, these effects can be properly described within the RG approach,
but one should combine the RG and OPE techniques and go beyond the plain
$\eps$ expansions. Analysis of the $L$ dependence of the Galilean invariant
objects like the structure functions requires the explicit construction of
all dangerous invariant scalar operators, exact calculation of their critical
dimensions, and summation of their contributions in the corresponding OPE.
This is clearly not a simple problem and it requires considerable improvement
of the present techniques.

As the intermediate step in the investigation of intermittency and anomalous
scaling it is important to study simplified models that are, in a number of
respects, closer to the real NS turbulence but still allow for analytical
treatment. An important step is breaking the artificial assumption of
the time decorrelation of the advecting velocity field; see the remarks
in Refs. \cite{CV,SCL}.

In Refs. \cite{RG3,Juha2} (see also \cite{RG4} for the case of compressible
flow) the RG and OPE were applied to the problem of a passive scalar advected
by a Gaussian self-similar velocity with finite (and not small) correlation
time. The energy spectrum of the velocity in the inertial range has the form
$E(k)\propto k^{1-2\eps}$, while the correlation time at the momentum $k$
scales as $k^{-2+\eta}$. It was shown that, depending on the values of the
exponents $\eps$ and $\eta$, the model reveals various types of
inertial-range scaling regimes with nontrivial anomalous exponents, which
were explicitly derived to the first \cite{RG3,RG4} and second \cite{Juha2}
orders of the double expansion in $\eps$ and $\eta$. Earlier, a similar
model was proposed and studied in detail (using numerical simulations,
in two dimensions) in \cite{OU}. Various aspects of the transport and
dispersion of particles in random Gaussian self-similar velocity fields with
finite correlation time were also studied in Refs. \cite{AM,Fann,Horvai}.

Another important step toward the NS turbulence is to consider the turbulent
advection of passive {\it vector} fields. The latter can have different
physical meaning: magnetic field in the Kazantsev-Kraichnan model of
hydromagnetic turbulence in the kinematic approximation; perturbation in the
linearized NS equation with prescribed statistics of the background field;
density of an impurity with internal degrees of freedom, etc.

Despite the obvious practical significance of these physical situations,
the passive vector problem is especially interesting because of the
insight it offers into the inertial-range behavior of the NS turbulence.
Owing to the coupling between different
components of the vector field (both by the dynamical equation and the
incompressibility condition) and to the presence of a new stretching term in
the dynamical equation, that couples the advected quantity to the gradient
of the advecting velocity, the behavior of the passive vector field appears
much richer than that of the scalar field: ``...there is considerably more
life in the large-scale transport of vector quantities,'' (p. 232 of Ref.
\cite{Legacy}). Indeed, passive vector fields reveal anomalous scaling
already on the level of the pair correlation function \cite{V96}.
They also develop interesting large-scale instabilities that can be
interpreted as manifestation of the dynamo effect (in kinematic
approximation); see e.g. \cite{V96,KZ,DV}. Other important issues are
mixing of composite operators, responsible for the anomalous scaling,
and the effects of pressure on the inertial-range behavior, especially
in anisotropic sectors.

In the scalar case, the anomalous exponents for all structure functions
are given by a single expression which includes $n$, the order of a
function, as a parameter \cite{Falk1,GK,RG}. This remains true for the
vector models with the stretching term \cite{Lanotte2,amodel}. In the
special vector model without the stretching term, considered e.g. in
\cite{LA,vector}, the number and the form of the operators
entering into the relevant family depend essentially on $n$, and different
structure functions should be studied separately. As a result, no general
expression valid for all $n$ exists in the model, and the anomalous exponents
are related by finite {\it families} of composite operators rather than by
individual operators \cite{LA,vector}. In this respect, such a model is
closer to the nonlinear NS equation, where the inertial-range behavior of
structure functions is believed to be related with the Galilean-invariant
operators, which form infinite families that mix heavily in renormalization;
see \cite{UFN,turbo}.

Another important question that can be addressed for the passive vector
model is the effects of nonlocal {\it pressure} terms on the anomalous
scaling, in particular, the consistency of the hierarchical picture
for the anisotropic anomalous contributions, known for the pressureless
scalar \cite{RG3,RG4,ST} and magnetic \cite{Lanotte,Lanotte2} cases,
with the presence of nonlocal terms in the closed equations for the
correlation functions, caused by the pressure contributions \cite{AP}
(a more detailed treatment is given in \cite{vector}).

The general vector model, introduced in \cite{amodel}, includes as special
cases the kinematic magnetic model, linearized NS equation and the special
model without the stretching term, and thus allows one to control the
pressure contribution and quantitatively study its effects on the anomalous
scaling. The generalized model also naturally arises within the multiscale
technique, as a result of the vertex renormalization \cite{Legacy}.

Finally, it should be noted that, as the experience with the passive fields
shows, the stochastic NS equation will show anomalous scaling already in its
linearized form. Thus the results obtained for the passive vector quantity,
with an appropriate statistics of the background field, can be considered as
an approximation to the full-fledged NS problem, which, in principle, can be
systematically improved by including nonlinear term as the perturbation.
It was also argued that, with a proper choice of the random stirring,
the passive vector model can reproduce the anomalous exponents of the NS
velocity field exactly \cite{BBT}.

The aforementioned works, however, have mostly been confined with the
Kraichnan velocity ensemble with zero correlation time.

In the present paper, we consider the model of the passive vector field
with the most general form of the nonlinear term,
advected by the synthetic velocity ensemble with a finite (and
not small) correlation time. The advected velocity field is Gaussian,
with the inertial-range spectrum of the form $E(k)\propto k^{1-2\eps}$
and the dispersion law $\omega\propto k^{2-\eta}$, where $k$ is the
momentum (wave number). Thus we generalize the general vector model
studied in \cite{amodel} for the zero-correlated case, to the ensemble
of the advecting velocity field employed, e.g., in  Refs.
\cite{RG3,Juha2,RG4} for the case of passive scalar.

We shall study anomalous scaling, stability of the scaling regimes and
analytically derive the anomalous exponents to the first order in
$\eps\sim\eta$. This allows us to investigate the universality of the
anomalous exponents and the effects of compressibility, pressure,
correlation time and large-scale anisotropy on the inertial-range anomalous
scaling. The plan of the paper is as follows.

In Sec.~\ref{sec:FT} we give detailed definition of the general vector
model and the advecting velocity ensemble and discuss its interesting
special cases: the rapid-change and frozen regimes, kinematic dynamo model
and linearized NS equation, and so on. We give the field theoretic
formulation of the original stochastic problem and present the corresponding
diagrammatic technique. In Sec.~\ref{sec:RG} we analyze the UV divergences
of the model, establish its multiplicative renormalizability and present
the renormalization constants in the one-loop approximation.
In Sec.~\ref{sec:Fixed} we analyze possible scaling regimes of the model,
associated with nontrivial and physically acceptable fixed points of the
corresponding RG equations. There are seven such regimes, any one of them
can be realized depending of the values of the model parameters
($\eps$, $\eta$ and others).
We discuss the physical meaning of these regimes (e.g., some of them
correspond to zero, finite, or infinite correlation time of the
advecting field, to the magnetic case or linearized NS equation)
and their regions of stability in the space of the model parameters.
In Sec.~\ref{sec:Scaling} we give the scaling representation for a
general correlation function and present the general expression for
the critical dimension of a composite field (operator). Then we consider
the most interesting case of tensor composite operators built only of
the passive vector field, which play a crucial role in further discussion
of the anomalous scaling. The critical dimension of such operator with
arbitrary number of the fields and vector indices is presented to first
order in $\eps\sim\eta$. In Sec.~\ref{sec:OPE} we introduce the
operator-product expansion and demonstrate its relevance to the issue of
inertial-range anomalous scaling. We show that the anomalous exponents in
our model can be identified with the critical dimensions of aforementioned
composite operators and present the leading terms of the inertial-range
behavior for a number of correlation functions. In Sec.~\ref{sec:Hier}
we show that the anomalous exponents of anisotropic contributions, determined
by the critical dimensions of tensor composite fields, obey the hierarchy
relations similar to those known for the passive scalar and zero-correlated
cases. We also discuss the dependence of the anomalous exponents
on the compressibility and pressure, and the effects of these factors on the
hierarchy relations. The results obtained are briefly reviewed and discussed
in the Conclusion.

\section{Description of the model. The field theoretic formulation}
\label{sec:FT}

Here and below, we denote $x \equiv \{t, {\bf x}\}$,
$\partial_{t} \equiv \partial /\partial t$,
$\partial _i \equiv \partial /\partial x_{i}$, and $d$ is the
(arbitrary) dimensionality of the ${\bf x}$ space.

We confine ourselves to the case of transverse (divergence-free)
passive vector field $\theta_{i}(x)$,
while the advecting field ${\bf v}(x)\equiv \{v_{i}(x)\}$ may
have a longitudinal (potential) component, so that
$\partial_i \theta_i=0$ and
$\partial_i v_i \ne 0$.
Thus the advection-diffusion equation has the form
\begin{eqnarray}
\partial_t \theta_i +  V_i^{(1)} - {\cal A}_0 V_i^{(2)} +
\partial_i {\cal P} = \kappa_0 \partial^2 \theta_i + f_i,
\label{1}
\end{eqnarray}
with the nonlinear terms
\begin{eqnarray}
V_i^{(1)} \equiv \partial_j \left(v_j \theta_i\right), \qquad
V_i^{(2)} \equiv \partial_j \left(v_i \theta_j\right) =
\theta_j \partial_j v_i.
\label{vertexes}
\end{eqnarray}
Here $\A_0$ is an arbitrary parameter, ${\cal P}(x)$ is the pressure,
$\kappa_0$ is the diffusivity, $\partial^2$ is the Laplace operator and
$f_{i}(x)$ is a Gaussian stirring force with zero mean and correlator
\begin{equation}
\langle  f_{i}(x)  f_{j}(x')\rangle
= \delta(t-t')\, C_{ij}({\bf r}/L), \quad
{\bf r}={\bf x}-{\bf x'}.
\label{2}
\end{equation}
The parameter $L$ is an integral scale related to the stirring, and $C_{ij}$
is a dimensionless function finite as $L\to\infty$. Its precise form is
unessential; for generality, it is not assumed to be isotropic. The force
$f_{i}(x)$ maintains the steady state of the system and gives rise to nonzero
correlation functions of the field $\theta$. In a more realistic formulation
it is replaced by an imposed nonzero mean value $\langle\theta\rangle$, see
e.g. \cite{Lanotte,Lanotte2}.

Nonlinear terms are chosen in the form of total derivatives, so that equation
(\ref{1}) is the conservation law for $\theta$ and, for $\A_{0}=1$, gives the
well-known equation for the magnetic field in the hydromagnetic
problem. The amplitude factor in front of the first nonlinear term in
(\ref{1}) can be absorbed by rescaling of the velocity field and
thus we set it to unity. The third possible structure,
$V_i^{(3)} \equiv \partial_i \left(v_j \theta_j\right)$,
can be absorbed into the pressure term $\partial_i {\cal P}$.

Besides the magnetic case ($\A_{0}=1$), the
model (\ref{1}) includes as special cases the linearized NS
equation with prescribed statistics of the background field
($\A_{0}=-1$), and the model of passively advected vector impurity
($\A_{0}=0$), which possesses an additional symmetry,
$\theta\to\theta+{\rm const}$,
and has an intrinsic formal resemblance with the stochastic
NS equation; see \cite{LA}. In these examples, the vector field has
different physical
interpretations: magnetic field, weak perturbation of the prescribed
background flow, concentration or density of the impurity particles
with an internal structure.

Owing to the transversality condition, the pressure can be
expressed as the solution of the Poisson equation,
\begin{equation}
 \partial^{2}{\cal P} = (\A_{0}-1)\, \partial_{i} \partial_{j}
(v_{j} \theta_{i}) .
\label{Pressure}
\end{equation}
For $\A_{0}=1$ (magnetic case) the pressure vanishes.
In this case Eq. (\ref{1}) also describes dynamics of the vorticity field
advected by a given background velocity field, see e.g. \cite{Legacy}.

In the real problem, the velocity ${\bf v}(x)$ satisfies the NS
equation, probably with additional terms that describe the feedback
of the advected field $\theta(x)$. We shall begin, however, with a
simplified model where the statistics of ${\bf v}(x)$ is given: it
is a Gaussian field with zero mean and correlation function
\begin{equation}
\big\langle v_{i}(x) v_{j}(x')\big\rangle =
\int \frac{d\omega}{2\pi}  \int \frac{d{\bf k}}{(2\pi)^d}
\Bigl\{ P_{ij}(\k) + \alpha Q_{ij}(\k) \Bigr\}
\, D_{v}(\omega,k)
\exp \left[ -{\rm i} (t-t')+{\rm i}{\bf k}\cdot({\bf x}-{\bf x'})\right] .
\label{f}
\end{equation}
Here $P_{ij}({\bf k}) = \delta _{ij} - k_i k_j / k^2$ and
$Q_{ij}({\bf k}) = k_i k_j / k^2$ are the transverse and the
longitudinal projectors, respectively, $d$ is the dimensionality
of the ${\bf x}$ space, $\alpha\ge0$ is a free parameter
($\alpha=0$ corresponds to the divergence-free advecting field,
$\partial_i v_i=0$). For the function $D_{v}$ we choose
\begin{equation}
D_{v}(\omega,k)=  \frac{g_{0}u_{0}\kappa_0^{3}\, {k}^{4-d-2\eps-\eta}}
{\omega^{2}+[u_{0}\kappa_0\, {k}^{2-\eta}]^{2}}\, .
\label{Fin}
\end{equation}
For the energy spectrum of the field $v$ we thus obtain
$E(k) \simeq k^{d-1} \int d\omega D_{v}(\omega,k)
\simeq g_{0} \kappa_0^{2} \, k^{1-2\eps}$. Therefore, the coupling
constant $g_{0}>0$ and the exponent $\eps$ describe
the equal-time velocity correlator or, equivalently, the energy spectrum,
while the constant $u_{0}>0$ and the exponent $\eta$
are related to the frequency $\omega\simeq u_{0}\kappa_0\, {k}^{2-\eta}$
characteristic of the mode $k$. The factor $\kappa_0^{3}$ in the
numerator of Eq. (\ref{Fin}) is explicitly isolated for the convenience
later on.

The exponents $\eps$ and $\eta$ are the analogs of the RG
expansion parameter $\eps=4-d$ in the theory of critical behavior,
and we shall use the traditional term  ``$\eps$  expansion'' for
the double expansion in the $\eps$--$\eta$ plane around the origin
$\eps=\eta=0$, with the additional convention that $\eta=O(\eps)$.
The IR regularization is provided by the cut-off in the integral
(\ref{f}) from below at $k\simeq m$, where $m\sim 1/L$ is the
reciprocal of the integral scale. Dimensionality considerations
show that the coupling constants $g_{0}$, $u_{0}$ are related to
the characteristic UV momentum scale $\Lambda \sim 1/\ell$ by
\begin{equation}
g_{0}\simeq \Lambda^{2\eps}, \quad u_{0}\simeq \Lambda^{\eta}.
\label{gg}
\end{equation}

The model (\ref{f}), (\ref{Fin}) contains two special cases that
possess some interest on their own. In the limit
$u_{0}\to\infty$, $g_{0}'\equiv g_{0}/u_{0}=\const$
we arrive at the rapid-change model:
\begin{equation}
D_{v}(\omega,k)\to g_{0}'\kappa_0\, k^{-d-\zeta},
\quad \zeta\equiv 2\eps - \eta,
\label{RC1}
\end{equation}
and the limit $u_{0}\to 0$, $g_{0}=\const$ corresponds to the case
of a ``frozen'' velocity field (or ``quenched disorder''):
\begin{equation}
D_{v}(\omega,k)\to g_{0}\kappa_0^{2}\, k^{-d+2-2\eps}\, \pi\,\delta(\omega),
\label{RC2}
\end{equation}
then the velocity correlator (\ref{f}) is independent of the time variable
$t-t'$ in the $t$ representation.

The stochastic problem (\ref{1}), (\ref{2}), (\ref{f}) is equivalent to
the field theoretic model of the extended set of three fields
$\Phi\equiv\{\theta', \theta, {\bf v}\}$ with action functional
\begin{equation}
S(\Phi)=\theta' D_{f}\theta' /2+
\theta' \left[ - \partial_{t} + \kappa _0\partial^{2}\theta
-  V^{(1)} + {\cal A}_0 V^{(2)} \right] -
{\bf v} D_{v}^{-1} {\bf v}/2
\label{action}
\end{equation}
with $V^{(1,2)}$ from (\ref{vertexes}). This means that statistical averages
of random quantities in the original stochastic problem can be represented
as functional averages with the weight $\exp S(\Phi)$.
The first five terms in Eq. (\ref{action}) represent the so-called
Martin--Siggia--Rose action for the stochastic problem (\ref{1}),
(\ref{2}) at fixed ${\bf v}$ (see, e.g., \cite{book,UFN,turbo} and references
therein), while the last term represents the Gaussian averaging over
${\bf v}$. Here $D_{f}$ and $D_{v}$ are the correlation functions
(\ref{2}) and (\ref{f}), respectively, $\theta'\equiv\theta'_i(x)$
is an auxiliary transverse
vector field, the required integrations over $x=(t,{\bf x})$ and summations
over the vector indices are implied, for example,
\[ \theta' \partial_{t} \theta \equiv \int dt\, d{\bf x}\, \theta'_{i}
(t,{\bf x})\, \partial_{t}\, \theta_{i} (t,{\bf x}). \]
The pressure term can be omitted in the functional (\ref{action})
owing to the transversality of the auxiliary field:
\[ \int dx \theta_{i}' \partial_{i} {\cal P} =
- \int dx {\cal P} \partial_{i} \theta_{i}' =0 .\]
Of course, this does not mean that the pressure contribution can simply be
neglected: the field $\theta'$ acts as the transverse projector and
selects the transverse part of the expression in the square brackets in
Eq.~(\ref{action}).

The model (\ref{action}) corresponds to a standard Feynman
diagrammatic technique with the triple vertex
$\theta' [-  V^{(1)} + {\cal A}_0 V^{(2)} ]$
and bare propagators in the frequency--momentum
($\omega$--$\k$) representation
\begin{eqnarray}
\bigl\langle \theta_{i} \theta'_{j} \bigr\rangle _0 =
\frac{P_{ij}(\k)} {(-{\rm i}\omega +\kappa _0 k^2)}, \quad
\bigl\langle \theta_{i} \theta_{j}\bigr\rangle _0
= \frac {C_{ij}(\k)} {(\omega^{2}+\kappa _0 ^2 k^{4})}, \quad
\bigl\langle \theta_{i}' \theta_{j}' \bigr\rangle _0 = 0,
\label{lines}
\end{eqnarray}
where $C_{ij}(\k)$ is the Fourier transform of the function
$C_{ij}({\bf r}/L)$ from (\ref{2}); the bare propagator
$\langle v_{i} v_{j}  \rangle _0$ is given by Eq.~(\ref{f}).

\section{UV renormalization. RG functions and RG equations}
\label{sec:RG}

The analysis of UV divergences is based on the analysis of
canonical dimensions. Dynamical models
of the type (\ref{action}), in contrast to static models, have
two scales, i.e., the canonical dimension of
some quantity $F$ (a field or a parameter in the action functional)
is described by two numbers, the momentum dimension $d_{F}^{k}$ and
the frequency dimension $d_{F}^{\omega}$; see e.g. \cite{book,UFN,turbo}.
They are determined so that
$[F] \sim [L]^{-d_{F}^{k}} [T]^{-d_{F}^{\omega}}$, where $L$ is the
length scale and $T$ is the time scale. The dimensions are found
from the obvious
normalization conditions $d_k^k=-d_{\bf x}^k=1$, $d_k^{\omega }
=d_{\bf x}^{\omega }=0$, $d_{\omega }^k=d_t^k=0$,
$d_{\omega }^{\omega }=-d_t^{\omega }=1$, and from the requirement
that each term of the action functional be dimensionless (with
respect to the momentum and frequency dimensions separately).
Then, based on $d_{F}^{k}$ and $d_{F}^{\omega}$,
one can introduce the total canonical dimension
$d_{F}=d_{F}^{k}+2d_{F}^{\omega}$ (in the free theory,
$\partial_{t}\propto\partial^{2}$), which plays in the theory of
renormalization of dynamical models the same role as
the conventional (momentum) dimension does in static problems.

The dimensions for the model (\ref{action}) are given in Table~\ref{table1},
including the parameters which will be introduced later on.
From the Table it follows that the model is
logarithmic (the coupling constants $g_{0}$, $u_{0}$ are dimensionless)
at $\eps=\eta=0$, so that the UV divergences in the correlation functions
have the form of the poles in $\eps$, $\eta$ and their linear combinations.

The total canonical dimension of an arbitrary
1-irreducible correlation function $\Gamma = \langle\Phi \cdots \Phi
\rangle _{\rm 1-ir}$ is given by the relation
$d_{\Gamma }=d_{\Gamma }^k+2d_{\Gamma }^{\omega }=
d+2-N_{\Phi }d_{\Phi}$,
where $N_{\Phi}=\{N_{\theta},N_{\theta'},N_{\bf v}\}$ are the
numbers of corresponding fields entering into the function
$\Gamma$, and the summation over all types of the fields is
implied. The total dimension $d_{\Gamma}$ is the formal index of the
UV divergence. Superficial UV divergences, whose removal requires
counterterms, can be present only in those functions $\Gamma$ for
which $d_{\Gamma}$ is a non-negative integer.

Analysis of the divergences in our model can be augmented by
the following considerations:

(i) From the explicit form of the vertex and bare propagators
it follows that $N_{\theta'}- N_{\theta}=2N_{0}$
for any 1-irreducible correlation function, where $N_{0}\ge0$
is the total number of bare propagators $\langle \theta \theta
\rangle _0$ entering into the function.
Therefore, the difference
$N_{\theta'}- N_{\theta}$ is an even non-negative integer for
any nonvanishing function; cf. Refs. \cite{RG,RG3,RG4}.

(ii) For any model with the Martin--Siggia--Rose-type action, all
the 1-irreducible functions with $N_{\theta'}=0$ contain closed
contours of retarded propagators $\langle\theta\theta'\rangle_{0}$
and vanish; see e.g. \cite{book,UFN,turbo}.

(iii) If for some reason a number of external momenta occurs as an
overall factor in all the diagrams of a given Green function, the
real index of divergence $d_{\Gamma}'$ is smaller than $d_{\Gamma}$
by the corresponding number (the correlation function requires
counterterms only if $d_{\Gamma}'$  is a non-negative integer);
see e.g. \cite{book,UFN,turbo}.
In the model (\ref{action}), the derivative $\partial$ at the
vertex $\theta' [-  V^{(1)} + {\cal A}_0 V^{(2)} ]$
can be moved onto the field $\theta'$ using the integration by parts,
which decreases the real index of divergence:
$d_{\Gamma}' = d_{\Gamma}- N_{\theta'}$, and the field $\theta'$ enters
the counterterms only in the form of the derivative $\partial\theta'$.

From the dimensions in Table~\ref{table1} we find $d_{\Gamma} =
d+2 - N_{\bf v} + N_{\theta}- (d+1)N_{\theta'}$ and $d_{\Gamma}'=
(d+2)(1-N_{\theta'}) + N_{\theta} - N_{\bf v}$. Bearing in mind
that $N_{\theta'}\ge N_{\theta}$ we conclude that for any $d$,
superficial divergences can exist only in the 1-irreducible
functions $\langle\theta'\rangle_{\rm 1-ir}$ with $d_{\Gamma}
=d_{\Gamma}'=1$, $\langle\theta'\theta\rangle_{\rm 1-ir}$ with
$d_{\Gamma} =2$, $d_{\Gamma}'=1$ and $\langle\theta'\theta{\bf
v}\rangle_{\rm 1-ir}$ with $d_{\Gamma} =1$, $d_{\Gamma}'=0$. The
corresponding counterterms necessarily reduce to the forms
$\partial\theta'$ (which vanishes identically),
$\theta'\partial^{2}\theta$,  $\theta'V^{(1)}$ and
$\theta'V^{(2)}$ with $V^{(1,2)}$ from (\ref{vertexes}). The
structure $\theta'\partial_{t}\theta$ does not contain a spatial
derivative, while $\theta'V^{(3)}$ with $V_i^{(3)} \equiv
\partial_i \left(v_j \theta_j\right)$ has the form of a total
derivative and vanishes after the integration over ${\bf x}$.

We thus conclude that our model (\ref{action}) is multiplicatively
renormalizable and the corresponding renormalized action has the form
\begin{equation}
S_{R}(\Phi)=
\theta' D_{f}\theta' /2+
\theta' \left[ - \partial_{t} + \kappa Z_{\kappa}  \partial^{2} \theta
- Z_{1}  V^{(1)} + Z_{2} {\cal A} V^{(2)} \right]
-{\bf v} D_{v}^{-1} {\bf v}/2.
\label{Rac}
\end{equation}
Here and below $g$, $u$, $\kappa$ and $\A$ (without a subscript) denote
the renormalized analogs of the corresponding bare parameters (with
the subscript 0). The correlation function $D_{v}$ in (\ref{Rac})
should be expressed in terms of the renormalized parameters and
$Z_{i}=Z_{i}(g,u,\A,\alpha,\eps,\eta,d)$ are the renormalization constants.
The introduction of the counterterms is reproduced by the multiplicative
renormalization of the velocity field, ${\bf v}\to Z_{1}{\bf v}$, and
the parameters $g_{0}$, $u_{0}$, $\kappa_0$ and $\A_{0}$ in the action
functional (\ref{action}):
\begin{equation}
\kappa_{0}=\kappa Z_{\kappa},  \quad u_{0}=u\mu^{\eta}\, Z_{u}
\quad g_{0}=g\mu^{2\eps}\, Z_{g}, \quad \A_{0} = Z_{\A} \A.
\label{mult}
\end{equation}
Here $\mu$ is the reference mass (additional arbitrary parameter of the
renormalized theory) and the renormalization constants in Eqs. (\ref{Rac})
and (\ref{mult}) are related as follows:
\begin{equation}
Z_{g}= Z_{1}^{2} Z_{\kappa}^{-2}, \quad
Z_{u}= Z_{\kappa}^{-1}, \quad
Z_{\A}= Z_{2} Z_{1}^{-1}.
\label{mult2}
\end{equation}
The first two relations in Eq. (\ref{mult2}) result from the absence
of the renormalization of the term with $D_{v}$ in (\ref{Rac}).
No renormalization of the fields $\theta$, $\theta'$ and the
parameters $m\sim 1/L$ and $\alpha$ is required, i.e., $Z_{\theta}=1$
and so on.

We have calculated all the renormalization constants in the one-loop
approximation (first order in $g$). The resulting expressions are
rather cumbersome: they are given by infinite series in the parameter
$u$ with the terms containing the poles in
$2\eps+ s \eta$ with $s=1,2,\dots$,
cf. Refs. \cite{RG3,RG4} for the scalar case. For this reason, below we
give them only for the special case $\eta=0$ and arbitrary $\eps$.
It is important here that the parameter $\eps$ alone provides the
UV regularization for the theory, so that the constants $Z$ remain finite
at $\eta=0$. In the minimal subtraction (MS) scheme they have the form:
\begin{mathletters}
\label{Z}
\begin{eqnarray}
Z_{1}= 1+ \frac{g\bar S_{d}} {4d(1+u)^{2}\eps} \left\{
\alpha + \frac{{\cal A}(1-{\cal A})}{(d+2)} -\alpha
\frac{(1-{\cal A})}{(d+2)} \right\} + O(g^{2}),
\label{Z1}
\end{eqnarray}
\begin{eqnarray}
Z_{2}= 1+ \frac{g\bar S_{d}} {4d(1+u)^{2}\eps} \left\{
\alpha - \frac{(1-{\cal A})}{(d+2)} +\alpha
\frac{(1-{\cal A})}{{\cal A}(d+2)} \right\}+ O(g^{2}),
\label{Z2}
\end{eqnarray}
\begin{eqnarray}
Z_{\kappa}= 1- \frac{g\bar S_{d}} {4(1+u)\eps} \left\{
{\cal X}+(\alpha -1){\cal Y} \right\}+ O(g^{2}),
\label{Znu}
\end{eqnarray}
\end{mathletters}
where
\begin{mathletters}
\label{XY}
\begin{eqnarray}
{\cal X}=1-\frac{(1-{\cal A})^{2}}{d} -\frac{2}{d(u+1)} +
\frac{2(u+2)(1-{\cal A})^{2}} {(u+1)d(d+2)}.
\label{X}
\end{eqnarray}
\begin{eqnarray}
{\cal Y}=\frac{1+{\cal A}-{\cal A}^{2}}{d}+
\frac{(1-{\cal A})^{2}}{d(d+2)}-\frac{2}{d(u+1)}+
\frac{2(1-{\cal A})}{d(d+2)(u+1)}.
\label{Y}
\end{eqnarray}
\end{mathletters}
Here and below $\bar S_{d}=S_{d}/(2\pi)^{d}$ and
$S_d = 2\pi ^{d/2}/\Gamma (d/2)$
is the surface area of the unit sphere in $d$-dimensional space.

The explicit expressions (\ref{Z}) illustrate some general
properties of the renormalization constants $Z_{1,2}$,
valid to all orders in $g$.

For the magnetic case ${\cal A}=1$ we have $Z_{1}=Z_{2}$ and, therefore,
$Z_{\A}=1$ and the parameter $\A_{0}$ is not renormalized: $\A_{0}=\A=1$.
This is a consequence of the relation
$\partial_{i} \left[V_i^{(1)} - V_i^{(2)}\right]=0$
for the vectors (\ref{vertexes}), which expresses the transversality
of the vertex $\theta'_{i} \left[V_i^{(1)} - V_i^{(2)}\right]$ with
respect to the index of the field $\theta'$. The same property holds
for all the diagrams of the 1-irreducible function
$\langle \theta'\theta v \rangle_{\rm 1-ir}$ and, as a result, for
the corresponding counterterm, cf. Refs. \cite{MHD1,MHD2} for the case
of the active magnetic field interacting with the NS velocity field.

In the rapid-change limit ($u\to\infty$, $g/u^{2}={\rm const}$)
we obtain $Z_{1}=Z_{2}=1$ due to the fact that all the diagrams of
the function $\langle \theta'\theta v \rangle_{\rm 1-ir}$
contain effectively closed circuits of retarded
propagators $\langle\theta\theta'\rangle_{0}$ and therefore vanish;
it is crucial here that the correlation function (\ref{RC1})
is proportional to the $\delta$ function in time representation.
On the contrary, in the frozen limit ($u\to0$, $g/u={\rm const}$)
the constants $Z_{1,2}$ remain nontrivial.

One also obtains $Z_{1}=Z_{2}=1$ for $\A=\alpha=0$. In this case, the
derivative $\partial$ at the only vertex
$\theta'_{i} V_i^{(1)} \equiv \theta'_{i} ({\bf v}\partt)\theta_{i}$
can be moved onto either fields $\theta$ and $\theta'$, so that
the real index of divergence takes on the form
$d_{\Gamma}' = d_{\Gamma}- N_{\theta}- N_{\theta'}$
(we recall that $d_{\Gamma}' = d_{\Gamma}- N_{\theta'}$ for
general $\A$ and $\alpha$). This gives $d_{\Gamma}'=-1$ for
$N_{\theta}= N_{\theta'}=N_{v}=1$, so that
the function $\langle \theta'\theta v \rangle_{\rm 1-ir}$
is UV finite; cf. Ref. \cite{RG3} for the scalar case.
In other words, the counterterm to the vertex must
include two derivatives (one for the field $\theta$ and one for
$\theta'$), which is forbidden by the dimensionality considerations.

Finally, from Eqs. (\ref{Z}) it follows that $Z_{1,2}=1$ for
$\A=1$ and $\alpha=0$. We found no general explanation for this
relation, but checked that it remains true in the two-loop
approximation, so that we can only guarantee that $Z_{1,2}=1+O(g^3)$.

Let $W(e_{0})$ be some correlation function in the original model
(\ref{action}) and $W_{R}(e,\mu)$ its analog in the renormalized
theory with action (\ref{Rac}). Here $e_{0}$ is the complete set of
bare parameters, and $e$ is the set of their renormalized counterparts.
In the following, we shall not be interested in the correlation
functions involving the velocity field ${\bf v}$. Then the relation
$S(\theta, \theta', Z_{2}\,{\bf v}, e_{0})
=S_{R}(\theta, \theta', {\bf v},e,\mu)$ for the action functionals
yields $W(e_{0})=W_{R}(e,\mu)$
for any correlation function of the fields $\theta'$, $\theta$;
the only difference is in the choice of variables and in the form
of perturbation theory (in $g$ instead of $g_{0}$).

We use $\widetilde{\cal D}_{\mu}$ to denote the differential
operation $\mu\partial_{\mu}$ for fixed
$e_{0}$ and operate on both sides of this equation with it. This
gives the basic RG equation:
\begin{equation}
{\cal D}_{RG}\,W_{R}(e,\mu) = 0,
\label{RG1}
\end{equation}
where ${\cal D}_{RG}$ is the operation $\widetilde{\cal D}_{\mu}$
expressed in the renormalized variables:
\begin{equation}
{\cal D}_{RG}\equiv {\cal D}_{\mu} + \beta_{g}\partial_{g} +
\beta_{u}\partial_{u} + \beta_{\A}\partial_{\A}
-\gamma_{\kappa}{\cal D}_{\kappa}.
\label{RG2}
\end{equation}
In Eq. (\ref{RG2}), we have written ${\cal D}_{x}\equiv x\partial_{x}$ for
any variable $x$, the RG functions (the $\beta$ functions and
the anomalous dimensions $\gamma$) are defined as
\begin{equation}
\gamma_{i}\equiv \Dm \ln Z_{i}
\label{RGF1}
\end{equation}
for any renormalization constant $Z_{i}$ and
\begin{eqnarray}
\beta_{g}\equiv\Dm  g=2g[-\eps+\gamma_{\kappa}-\gamma_{1}], \quad
\beta_{u}\equiv\Dm  u=u[-\eta+\gamma_{\kappa}], \quad
\beta_{\A} \equiv\Dm \A=\A [\gamma_{1}-\gamma_{2}].
\label{betas}
\end{eqnarray}
The relations between $\beta$ and $\gamma$ in (\ref{betas})
result from the definitions and the relations (\ref{mult2}).

For the basis anomalous dimensions from the definitions and the
expressions (\ref{Z}) in the one-loop approximation we obtain:
\begin{mathletters}
\label{gammas}
\begin{equation}
\gamma_{1}= \frac{-g\bar S_{d}} {2d(1+u)^{2}} \left\{
\alpha + \frac{{\cal A}(1-{\cal A})}{(d+2)} -\alpha
\frac{(1-{\cal A})}{(d+2)} \right\} + O(g^{2}),
\label{gamma1}
\end{equation}
\begin{equation}
\gamma_{2}= \frac{-g\bar S_{d}} {2d(1+u)^{2}} \left\{
\alpha - \frac{(1-{\cal A})}{(d+2)} +\alpha
\frac{(1-{\cal A})}{{\cal A}(d+2)} \right\}+ O(g^{2}),
\label{gamma2}
\end{equation}
\begin{equation}
\gamma_{\kappa}= \frac{g\bar S_{d}} {2(1+u)}
\Bigl\{{\cal X}+(\alpha -1){\cal Y} \Bigr\}+ O(g^{2})
\label{gammanu}
\end{equation}
\end{mathletters}
with ${\cal X}$ and ${\cal Y}$ from (\ref{XY}).
It is also worth noting that the knowledge of the constants $Z$ at $\eta=0$
is in fact sufficient to calculate the $\beta$ functions (\ref{betas}) for
all $\eta$, $\eps$ because the anomalous dimensions (\ref{gammas}) are
independent of $\eta$, $\eps$, at least in the one-loop approximation.
This fact essentially simplified the two-loop calculation of the
anomalous dimensions for the scalar case performed in Ref. \cite{Juha2}.

\section{Fixed points and scaling regimes} \label{sec:Fixed}

It is well known that possible scaling regimes of a renormalizable model
are associated with IR attractive fixed points of the corresponding RG
equation; see e.g. Ref. \cite{book}. Roughly speaking, in solving the RG
equation all the renormalized coupling constants $g_{i}$ (that is, all
dimensionless parameters of the model) are replaced by the corresponding
invariant charges $\bar g_{i}(s)$,
where $s\equiv k/\mu$ in the momentum representation or $s\equiv 1/\mu r$
in the coordinate representation.
The invariant charges are determined as the solutions of the following
Cauchy problem
\begin{equation}
{\cal D}_{s} \bar g_{i}(s) = \beta_{i} \left(\{\bar g_{j}(s)\}\right),
\quad \bar g_{i}(1) = g_{i} \quad {\rm for \ all }\ i.
\label{RGI}
\end{equation}
Here $\bar g_{i}(s)$ is the full set of the invariant charges and
$\beta_{i} \equiv \Dm g_{i}$ are the corresponding $\beta$ functions.
In the IR asymptotic range ($s\to0$) the invariant charges tend to
IR attractive fixed points of the system (\ref{RGI}). The coordinates
$g_{i*}$ of the fixed points are found from the equations
$\beta_{i}(\{g_{j*}\})=0$ for all $i$.
The type of a fixed point is determined by the matrix
$\Omega=\{\Omega_{ij}=\partial\beta_{i}/\partial g_{j}\}$:
for an IR attractive fixed point the matrix $\Omega$ is positive definite, i.e.,
the real parts of all its eigenvalues are positive.

In our case, the coordinates $g_{*}$, $u_{*}$, $\A_{*}$ of the fixed
points are found from the equations
\begin{equation}
\beta_{g} (g_{*},u_{*},\A_{*})=\beta_{u} (g_{*},u_{*},\A_{*})=
\beta_{\A} (g_{*},u_{*},\A_{*}) =0
\label{points}
\end{equation}
with the $\beta$ functions given in Eq. (\ref{betas}). The coordinates
of the fixed points and the elements of the corresponding matrices
$\Omega$ depend on the remaining free parameters: $\eps$, $\eta$, $d$ and
$\alpha$.\footnote{Formally, $\alpha$ can be treated as the fourth
coupling constant. The corresponding $\beta$ function
$\beta_{\alpha}\equiv \Dm {\alpha}$ vanishes identically owing
to the fact that $\alpha$ is not renormalized. Therefore, the equation
$\beta_{\alpha}=0$ gives no additional constraint on the values
of the coupling constants at a fixed point.}

Below we list all possible fixed points of the system (\ref{points}),
giving their coordinates in the one-loop approximation, that is, to
first order in $\eps$ and $\eta$. We shall also present the inequalities
that determine the regions (in the space of parameters $\eps$, $\eta$, $d$
and $\alpha$) where those points are IR attractive. It should also be kept
in mind that admissible fixed points should satisfy the relations
$g_{*}\ge0$, $u_{*}\ge0$, which follow from the physical meaning ot these
parameters ($g$ is the amplitude of a pair correlation function and
$u$ is the ratio of the diffusivity and viscosity coefficients).

First of all, the trivial fixed point
\begin{mathletters}
\label{TR1}
\begin{equation}
g_{*}= u_{*}=0, \quad  \A_* \ {\rm arbitrary}
\label{triv1}
\end{equation}
should be mentioned. The corresponding matrix $\Omega$ is diagonal with
the diagonal elements (eigenvalues)
\begin{equation}
\lambda_1=0, \quad \lambda_2=-2\varepsilon, \quad \lambda_3=-\eta,
\end{equation}
\end{mathletters}
so that the point (\ref{triv1}) is IR attractive for $\eps<0$, $\eta<0$.
Since for $g=u=0$ all the three $\beta$ functions (\ref{betas}) vanish
simultaneously, the value of $\A$ at this fixed point remains arbitrary.
This degeneracy is reflected in the vanishing of $\lambda_1$.

We shall discuss the physical meaning of the point (\ref{triv1})
later and now turn to nontrivial
fixed points. Their analysis is simplified by the observation that the
one-loop function $\beta_{\A}$ factorizes into a part that depends only
on $\A$ and a part that depends only on $g$, $u$:
\begin{equation}
\beta_{\A} = \frac{g\bar S_d} {2d(1+u)^2}\, (\A^2-1)(\A-\alpha)+O(g^2),
\label{betaA}
\end{equation}
see Eq. (\ref{betas}), (\ref{gammas}).
Thus all possible values of $\A_*$ are found from the equation
$\beta_{\A} =0$ regardless of the values of the other couplings,
provided $g$ is nonzero and $g$, $u$ are not infinite:
\begin{equation}
\A_{*} = -1,\ 1,\ \alpha.
\label{A}
\end{equation}
Furthermore, the elements $\partial \beta_{\A}/ \partial g |_{\A=\A_{*}}
=\partial \beta_{\A}/ \partial u |_{\A=\A_{*}} =0 $ of the matrix $\Omega$
vanish for these values of $\A_*$ regardless of the values of $g$, $u$.
Thus the matrix $\Omega$ is block-triangular and one of its eigenvalues
coincides with the diagonal element $\Omega_{\A}\equiv
\partial \beta_{\A}/ \partial \A$. Its sign is completely determined by
the value of $\A_*$, so that we can check a necessary condition
$\Omega_{\A}>0$ for a fixed point to be IR attractive regardless of the
values of $g$, $u$: it is satisfied for $\A_{*} = -1$ for all $\alpha$,
for $\A_{*} = 1$ if $\alpha<1$, and for $\A_{*} = \alpha$ if $\alpha>1$.
This is easy to understand geometrically: $\Omega_{\A}>0$ for the
leftmost and rightmost points and $\Omega_{\A}<0$ for the point lying
in between them.

Therefore, for any $\alpha$ the condition $\Omega_{\A}>0$ is
simultaneously satisfied by two fixed points:
$\A_{*} = -1$ and $\A_{*} = {\rm max}\ \{1,\ \alpha\}$.
Factorization of the function (\ref{betaA}) allows one to analyze
the solution of the equation (\ref{RGI}) for the invariant charge
$\bar\A(s)$ independently of the remaining equations for the other
invariant charges: $\bar\A(s)$ will be attracted by the leftmost fixed
point $\A_{*} = -1$ if and only if the initial condition $\A=\bar\A(1)$
lies to the left of the unstable fixed point,
$\A< {\rm min}\ \{1,\ \alpha\}$, and by the rightmost point
$\A_{*} = {\rm max}\ \{1,\ \alpha\}$ if and only if
$\A> {\rm min}\ \{1,\ \alpha\}$.

Besides the finite values (\ref{A}), there is one more possibility that
formally corresponds to $\A_{*} =\infty$. More accurately it can be
revealed by the change of variables $a\equiv 1/\A$, $y\equiv g\A^2$;
then $a_{*} =0$ and $y_{*}$ is finite.
It describes the situation when the action (\ref{action}) contains
the only vertex $\theta'V^{(2)}$. Since in this vertex the derivative
can be moved onto either of the fields $\theta'$ and ${\bf v}$
[see Eq. (\ref{vertexes})], such a model is multiplicatively
renormalizable (the vertex $\theta'V^{(1)}$ is not generated by the
renormalization) and $Z_{2}=1$ identically. Thus the result $a_{*} =0$
is exact to all orders. However, one can easily check that the
eigenvalue of the matrix $\Omega$, equal to the diagonal element
$\Omega_{a}\equiv \partial \beta_{a}/ \partial a$, is always
negative, so that this point cannot be IR attractive. We shall
not discuss it in what follows.

To conclude the analysis of the equation $\beta_{\A}=0$, it remains
to note that the result $\A_{*} = 1$ in Eq. (\ref{A}) is exact
(a consequence of the relation $Z_{1}=Z_{2}$ for $\A=1$, see
Sec.~\ref{sec:RG}), while the other two can have corrections of
order $\eps \sim \eta$ and higher. The only exception is the result
$\A_{*} = 0$ for $\alpha=0$, which is also exact due to the relation
$Z_{1}=1$ for $\alpha=0$.

Substituting the values (\ref{A}) into the functions $\beta_{g}$,
$\beta_{u}$ from (\ref{betas}) and solving the equations
$\beta_{g}=\beta_{u}=0$ gives the values of the remaining coordinates
$g_{*}$ and $u_{*}$. For each value of $\A_{*}$, there are two solutions.
For the first of them, $u_{*}=0$. This result is exact to all orders
of the $\eps$ expansion since the function $\beta_{u}$ for $u=0$ vanishes
identically, see Eq. (\ref{betas}). Substituting the value $u_{*}=0$ to
$\beta_{g}$ and solving the equation $\beta_{g}=0$ gives $g_{*}$.
We shall denote such fixed points by $Q$, as they correspond to the
case of ``quenched disorder,''  see the comments to Eq. (\ref{RC2}).
For the second variant, $g_{*}$ and $u_{*}$ are both nonzero;
we shall denote such fixed points by $F$ (``finite'' correlation time
of the velocity field).

Thus we arrive at six nontrivial fixed points
$Q^{-}$, $Q^{+}$, $Q^{\alpha}$, $F^{-}$, $F^{+}$, $F^{\alpha}$,
where the superscripts correspond to the values of $\A_*$ in Eq. (\ref{A}).
Below we give the coordinates of these points in the one-loop approximation,
the eigenvalues $\lambda_{1,2,3}$ of the matrix $\Omega$ ($\lambda_{1}$
always corresponds to the diagonal element $\partial\beta_{\A}/\partial\A$),
and the inequalities that determine the regions where the points are
admissible (IR attractive and satisfy $g_{*}>0$, $u_{*} \ge 0$):

\begin{mathletters}
\label{FPQ+}
\begin{eqnarray}
Q^{+}: \qquad g_{*}=\frac{2 d \eps}{d-1}, \quad u_{*}=0, \quad \A_{*}=1
\label{Q+}
\end{eqnarray}
with the eigenvalues
\begin{eqnarray}
\lambda_1&=& \frac{2 \varepsilon(1-\alpha)}{(d-1)(d+2)},
\nonumber \\
\lambda_2&=& 2 \varepsilon,
\nonumber \\
\lambda_3&=&\frac{(d-1-\alpha) \varepsilon - \eta (d-1)}{d-1},
\label{100}
\end{eqnarray}
admissible for
\begin{eqnarray}
\alpha<1, \quad  (d-1-\alpha)\eps > (d-1) \eta.
\label{101}
\end{eqnarray}
\end{mathletters}

\begin{mathletters}
\label{FPQ-}
\begin{eqnarray}
Q^{-}: \qquad g_{*}= \frac{2d(d+2)\eps}{\bar S_{d}(d-2\alpha)(d-1)}\, ,
\quad u_{*}=0, \quad \A_{*}=-1
\label{Q-}
\end{eqnarray}
with the eigenvalues
\begin{eqnarray}
\lambda_1&=& \frac{2\eps (\alpha+1)} {(d-1)(d-2\alpha)},
\nonumber \\
\lambda_2&=& 2 \varepsilon,
\nonumber \\
\lambda_3 &=& \frac{ [-2+d-d^2+\alpha (3d-2)] \varepsilon-
(2 \alpha-d) (d-1) \eta}{(2 \alpha-d) (d-1)},
\label{102}
\end{eqnarray}
admissible for
\begin{eqnarray}
\eps>0, \quad \alpha<d/2, \quad
(d-2\alpha)(d-1)\eta <  [d^2-d+2 -\alpha(3d-2)] \eps.
\label{103}
\end{eqnarray}
\end{mathletters}

\begin{mathletters}
\label{FPQa}
\begin{eqnarray}
Q^{\alpha}: \qquad g_{*}=
\frac{ 2d(d+2) \varepsilon} {d^2-3+ \alpha (d+1)(1+\alpha-\alpha^2)},
\quad u_{*}=0, \quad \A_{*}=\alpha
\label{Qa}
\end{eqnarray}
with the eigenvalues
\begin{eqnarray}
\lambda_1&=& \frac{(\alpha^2-1) \varepsilon}{d^2+ \alpha (d+1)
(1+\alpha-\alpha^2)-3},
\nonumber \\
\lambda_2&=& 2 \varepsilon,
\nonumber \\
\lambda_3&=& \frac{[-3-\alpha+d^2+\alpha^2 (1+d)(1-\alpha)] \varepsilon
- [-3+d^2+\alpha (1+d)(1+\alpha-\alpha^2)] \eta}
{d^2+ \alpha (1+d)(1+\alpha-\alpha^2)-3},
\label{104}
\end{eqnarray}
admissible for
\begin{eqnarray}
\eps>0, \quad \alpha>1, \quad
d^2-3+\alpha(d+1)(1+\alpha-\alpha^2)>0,
\nonumber \\
\eta< \frac{d^2-3-\alpha+\alpha^2(d+1)(1-\alpha)}
{d^2-3+\alpha(d+1)(1+\alpha-\alpha^2)}\eps.
\label{105}
\end{eqnarray}
\end{mathletters}

\begin{mathletters}
\label{FPF+}
\begin{eqnarray}
F^{+}: \qquad g_{*}=
\frac{2 \alpha d (\eta -2 \varepsilon)^2}{(\alpha+ d -1)^2
(\varepsilon-\eta)},
\quad
u_{*}= \frac{(1+ \alpha - d) \varepsilon + (d-1) \eta}
{(d+\alpha -1) (\varepsilon-\eta)},
\quad \A_{*}=1
\label{F+}
\end{eqnarray}
with the eigenvalues
\begin{eqnarray}
\lambda_1&=& \frac{2 (1-\alpha) (\varepsilon-\eta)}{\alpha (d+2)},
\nonumber \\
\lambda_2&=& \frac{W_1-\sqrt{W_1^2-W_2}}{2 \alpha (2 \varepsilon - \eta)},
\nonumber \\
\lambda_3&=&\frac{W_1+\sqrt{W_1^2-W_2}}{2 \alpha (2 \varepsilon -\eta)},
\label{106}
\end{eqnarray}
where
\begin{eqnarray}
W_1&=& (2+6 \alpha-2 d) \varepsilon^2 + 5 (d-1-\alpha) \varepsilon
\eta - 3 (d-1) \eta^2, \nonumber \\
W_2&=& 8 \alpha (2 \varepsilon-\eta) ((1+\alpha-d)
\varepsilon+(d-1) \eta) (2 \varepsilon^2-3 \varepsilon
\eta+\eta^2),
\label{1107}
\end{eqnarray}
admissible for
\begin{eqnarray}
\varepsilon>0, \quad \eta>0, \quad \alpha<1, \quad
\varepsilon > \eta > \frac{d-\alpha-1}{d-1} \varepsilon,
\nonumber \\
(2+6 \alpha-2 d) \varepsilon^2 + 5 (d-1-\alpha)
\varepsilon \eta - 3 (d-1) \eta^2>0.
\label{1108}
\end{eqnarray}
\end{mathletters}

\begin{mathletters}
\label{FPF-}
\begin{eqnarray}
F^{-}: \qquad g_{*}=
\frac{2 d (2+d) (\alpha d - 2) (\eta -2 \varepsilon)^2
}{(1-\alpha +d)^2 (d-2)^2 (\varepsilon-\eta)},
\nonumber \\
u_{*}=
\frac{(2+ \alpha(2-3d) - d + d^2) \eps +
(2 \alpha-d) (d-1) \eta}{(\alpha -d-1) (d-2) (\varepsilon-\eta)},
\quad \A_{*}=-1
\label{F-}
\end{eqnarray}
with the eigenvalues
\begin{eqnarray}
\lambda_1&=& \frac{2 (1+\alpha) (\varepsilon-\eta)}{-2+\alpha d},
\nonumber \\
\lambda_2&=& \frac{W_1-\sqrt{W_1^2-W_2}}{2 (-2+\alpha d)
(2 \varepsilon-\eta)},
\nonumber \\
\lambda_3&=&\frac{W_1+\sqrt{W_1^2-W_2}}{2 (-2+\alpha d)
(2 \varepsilon-\eta)},
\label{107}
\end{eqnarray}
where
\begin{eqnarray}
W_1&=& 2 (-6+d-d^2+\alpha (-2+5 d)) \varepsilon^2 \nonumber \\
&+& 5 (2+\alpha (2-3 d)-d+d^2) \varepsilon
    \eta+3 (2 \alpha-d) (-1+d) \eta^2,
\nonumber \\
W_2&=& 8 (-2+\alpha d) (2 \varepsilon-\eta)^2 (\varepsilon-\eta)
\nonumber \\
&\times& ((-2+d-d^2+\alpha (-2+3 d)) \varepsilon-(2 \alpha-d)
(-1+d) \eta),
\label{108}
\end{eqnarray}
admissible for
\begin{eqnarray}
\eps>\eta, \quad  2/d<\alpha<(d+1), \quad
2 \varepsilon > \eta, \quad W_1 > 0,
\nonumber \\
(2+\alpha(2-3d)+d(d-1))\varepsilon<(d-1)(d-2\alpha)\eta.
\label{1109}
\end{eqnarray}
\end{mathletters}

\begin{mathletters}
\label{FPFa}
\begin{eqnarray}
F^{\alpha}: \qquad g_{*}=
\frac{2 \alpha d (2+d)^2 (\eta -2\varepsilon)^2 }{(d^2-3+\alpha^2
(1+d)(1-\alpha)+\alpha (2 d+3))^2 (\varepsilon-\eta)},
\quad  \A_{*}=\alpha,
\nonumber \\
u_{*}=
\frac{(3+\alpha-d^2-\alpha^2 (1+d)(1-\alpha)) \varepsilon +
(d^2-3 + \alpha (1+d)(1+\alpha-\alpha^2 )) \eta}{(d^2-3+\alpha^2
(1+d)(1-\alpha)+\alpha (2 d+3)) (\varepsilon-\eta)}
\label{Fa}
\end{eqnarray}
with the eigenvalues
\begin{eqnarray}
\lambda_1&=& \frac{(\alpha^2-1) (\varepsilon-\eta)}{\alpha (2+d)},
\nonumber \\
\lambda_2&=& \frac{W_1-\sqrt{W_1^2-W_2}}{2 \alpha (2+d)
(2 \varepsilon-\eta)}, \nonumber \\
\lambda_3&=&\frac{W_1+\sqrt{W_1^2-W_2}}{2 \alpha (2+d)
(2 \varepsilon-\eta)},
\label{109}
\end{eqnarray}
where
\begin{eqnarray}
W_1&=& 2 (3-d^2-\alpha^2 (1+d)+\alpha^3 (1+d)+\alpha (5+2 d)) \varepsilon^2
\nonumber \\
&-& 5 (3+\alpha-d^2-\alpha^2 (1+d)+\alpha^3 (1+d)) \varepsilon
    \eta+3 (3-d^2-\alpha (1+d)(1+\alpha -\alpha^2)) \eta^2,
\nonumber \\
W_2&=& 8 \alpha (2+d) (2 \varepsilon-\eta)^2 (\varepsilon -\eta)
\nonumber \\
&\times& ((3+\alpha-d^2-\alpha^2 (1+d)(1-\alpha)) \varepsilon
\nonumber \\
&& + (-3+d^2+\alpha (1+d)(1+\alpha-\alpha^2) \eta),
\label{110}
\end{eqnarray}
admissible for
\begin{eqnarray}
&{}&  \alpha>1, \quad \eps>\eta, \quad 2\eps>\eta, \quad W_{1}>0,
\nonumber \\
&{}&
(d^2-3+\alpha^2 (1+d)(1-\alpha)+\alpha (2 d+3)) >0,
\nonumber \\
&{}&
(3+\alpha-d^2-\alpha^2 (1+d)(1-\alpha))
\varepsilon + (d^2-3 + \alpha (1+d)(1+\alpha-\alpha^2 )) \eta>0.
\label{112}
\end{eqnarray}
\end{mathletters}

However, the above list is not exhaustive: besides the
fixed points with non-infinite values of $g_{*}$ and $u_{*}$, there
are points for which these parameters tend to infinity. They can be
revealed by the change of variables $x\equiv g/u$, $w\equiv 1/u$.
The corresponding $\beta$ functions are obtained by the chain rule:
\begin{eqnarray}
\beta_{x} \equiv \Dm x = (1/u) \beta_{g} -(g/u^{2}) \beta_{u}, \quad
\beta_{w} \equiv \Dm w = -(1/u^{2}) \beta_{u},
\label{betax}
\end{eqnarray}
with $\beta_{g,u}$ from (\ref{betas}) and the anomalous dimensions
(\ref{gammas}) expressed in the variables $x$, $w$. Solving the
equations $\beta_{x}=\beta_{w}=\beta_{\A}=0$ gives three fixed points
with finite values of $x_{*}$, $w_{*}$, which simply express the
points $F^{\pm,\, \alpha}$ in the new variables. Besides them,
there are two fixed points with $w_{*}=0$, left out in the analysis
performed in terms of $g$, $u$. It is clear from (\ref{RC1}) that the
choice $w=0$ corresponds to the rapid-change limit of our model. The
dimension $\gamma_{\kappa}$ from Eq. (\ref{gammas}) remains finite for
$w=0$, while $\gamma_{1}$ and $\gamma_{2}$ vanish. In fact, they vanish
to all orders of the perturbation theory in $x\propto g$ owing to the
exact relation $Z_{1}=Z_{2}=1$ that holds in the limit (\ref{RC1});
see the discussion in Sec.~\ref{sec:RG}. As a result, the function
$\beta_{A}$ vanishes identically for $w=0$ and the coordinate $\A_*$
remains arbitrary at such fixed points; see the remark below Eq.
(\ref{triv1}).

There is a trivial fixed point
\begin{mathletters}
\begin{equation}
x_{*}=w_{*}=0, \quad   \A_* \ {\rm arbitrary}
\label{triv2}
\end{equation}
with the eigenvalues
\begin{equation}
\lambda_1=0, \quad \lambda_2=\eta, \quad \lambda_3=\eta-2\eps,
\label{triv22}
\end{equation}
\end{mathletters}
and a non-trivial fixed point,
which we shall denote by $R^{\A}$ in what follows,
\begin{mathletters}
\begin{eqnarray}
R^{\A}: \qquad x_{*}&=&\frac{2 d (d+2) (2\varepsilon -\eta)} {\A
(1+\alpha) d+d^2+\alpha (3+d)-\A^2 (-1+\alpha+\alpha d)-3},
\nonumber\\
w_{*}&=&0, \quad \A_{*} \ {\rm arbitrary},
\label{RCP}
\end{eqnarray}
with the eigenvalues
\begin{eqnarray}
\lambda_1=0, \quad \lambda_2=-2(\varepsilon-\eta), \quad
\lambda_3=2\varepsilon-\eta,
\label{200}
\end{eqnarray}
admissible for
\begin{eqnarray}
\eta > \varepsilon > \eta/2,  \quad
\A (1+\alpha) d+d^2+\alpha (3+d)-\A^2
(-1+\alpha+\alpha d)-3>0.
\label{202}
\end{eqnarray}
\end{mathletters}
The notation $R^{\A}$ implies that this point corresponds to the
``rapid-change'' regime and that $\A_{*}=\A$ remains a free parameter.

Triviality of the points (\ref{triv1}) and (\ref{triv2}) implies the absence
of anomalous scaling; they correspond to diffusive-type regimes, for which
the convection (that is, the nonlinearity in Eq. (\ref{1})) can be treated
within ordinary perturbation theory and the standard methods of the
homogenization theory apply. More detailed discussion of such fixed points
(in particular, the difference between the ``quenched'' and ``rapid-change''
trivial fixed points) can be found Ref. \cite{RG3} for the example of the
passive scalar field, advected by the velocity ensemble (\ref{f}).

On the contrary, the nontrivial fixed points $Q^{\pm,\, \alpha}$,
$F^{\pm,\, \alpha}$ and $R^{\A}$ describe non-diffusive asymptotic
regimes, in which the competition of the diffusive and convective terms
in Eq. (\ref{1}) produces anomalous scaling behavior. The corresponding
anomalous exponents will be presented in the next Section, and now we
shall discuss the interplay between the possible scaling regimes.

Seven nontrivial IR attractive fixed points correspond to seven possible
scaling regimes; only one of them can be realized when the values of
all parameters $\alpha$, $\eps$, $\eta$, $d$ and $\A$ are given,
regardless of the values of the amplitudes $g$ and $u$ (see below).
In this sense, the asymptotic behavior in our model is universal,
and (as we shall see) the anomalous exponents depend only on the values
of the exponents $\eps$ and $\eta$ in the velocity correlation function
and on the parameters $\alpha$ (for all regimes) and $\A$ (for $R^{\A}$),
but they do not depend on the coupling constants $g$ and $u$.

Indeed, let us fix the value of $\A$. Then the solution of the RG equation
(\ref{RG1}) can be attracted by either of the following three sets: by the
set $\{F^{-}, Q^{-}, R^{\A}\}$ if $\A< {\rm min}\ \{1,\ \alpha\}$, by the
set $\{F^{+}, Q^{+}, R^{\A}\}$, if $\alpha<1$ and $\A > \alpha$, and by the
set $\{F^{\alpha}, Q^{\alpha}, R^{\A}\}$ if $\alpha>1$ and $\A > 1$.
(In all these cases, the value of $\A_*$ for the fixed point $R^{\A}$
in Eq. (\ref{RCP}) simply equals to $\A$).

For any of these three situations, only one fixed point in the list
$\{F, Q, R\}$ can be IR attractive for given $\eps$ and $\eta$.
Indeed, the analysis of the inequalities that determine the admissibility
regions for these points shows that these regions adjoin each other
without overlaps or gaps. The common boundary of the admissibility
regions for the points $F$ and $R$ is $\eps=\eta$
[one of the admissibility conditions for $R^{\A}$ is $\eps<\eta$,
see Eq. (\ref{202}), while one of the admissibility conditions for any
of the points $F^{\pm\, \alpha}$ is $\eps>\eta$, see Eqs. (\ref{1108}),
(\ref{1109}) and (\ref{112})].
The common boundary of the admissibility regions for the points
$F$ and $Q$ is $ (d-1-\alpha) \eps = (d-1) \eta$
for the pair $Q^{+}$, $F^{+}$ [see Eqs. (\ref{101}) and (\ref{1108})],
$({d^2-3+\alpha(d+1)(1+\alpha-\alpha^2)})\eta=
({d^2-3-\alpha+\alpha^2(d+1)(1-\alpha)}) \eps$
for the pair $Q^{-}$, $F^{-}$ [see Eqs. (\ref{103}) and (\ref{1109})]
and $(d^2-3+\alpha^2 (1+d)(1-\alpha)+\alpha (2 d+3)) =0$ for the pair
$F^{\alpha}$, $Q^{\alpha}$ [see Eqs. (\ref{105}) and (\ref{112})].

Therefore, for any given set of parameters $\alpha$, $\eps$, $\eta$, $d$
and $\A$, the RG trajectory can be attracted by the only one possible
nontrivial fixed point, regardless of the values of the parameters
$g$, $u$ [that is, the initial data for the Cauchy problem (\ref{RGI})].
If the trajectory is attracted by a trivial point, the model will show
diffusive-type behavior. Finally, if there is no IR attractive fixed point
for the given set of parameters, no definitive conclusion can be made about
the asymptotic behavior of the model within the framework of the $\eps$
expansion. In particular, if the trajectory comes to a region where the
parameters $g$, $u$ are negative, one can think that the steady state
of our system becomes unstable (by analogy with the RG theory of critical
phenomena, where such behavior is usually interpreted as a first-order
phase transition).

\begin{figure}
\centerline{ \hbox{ \epsfig{file=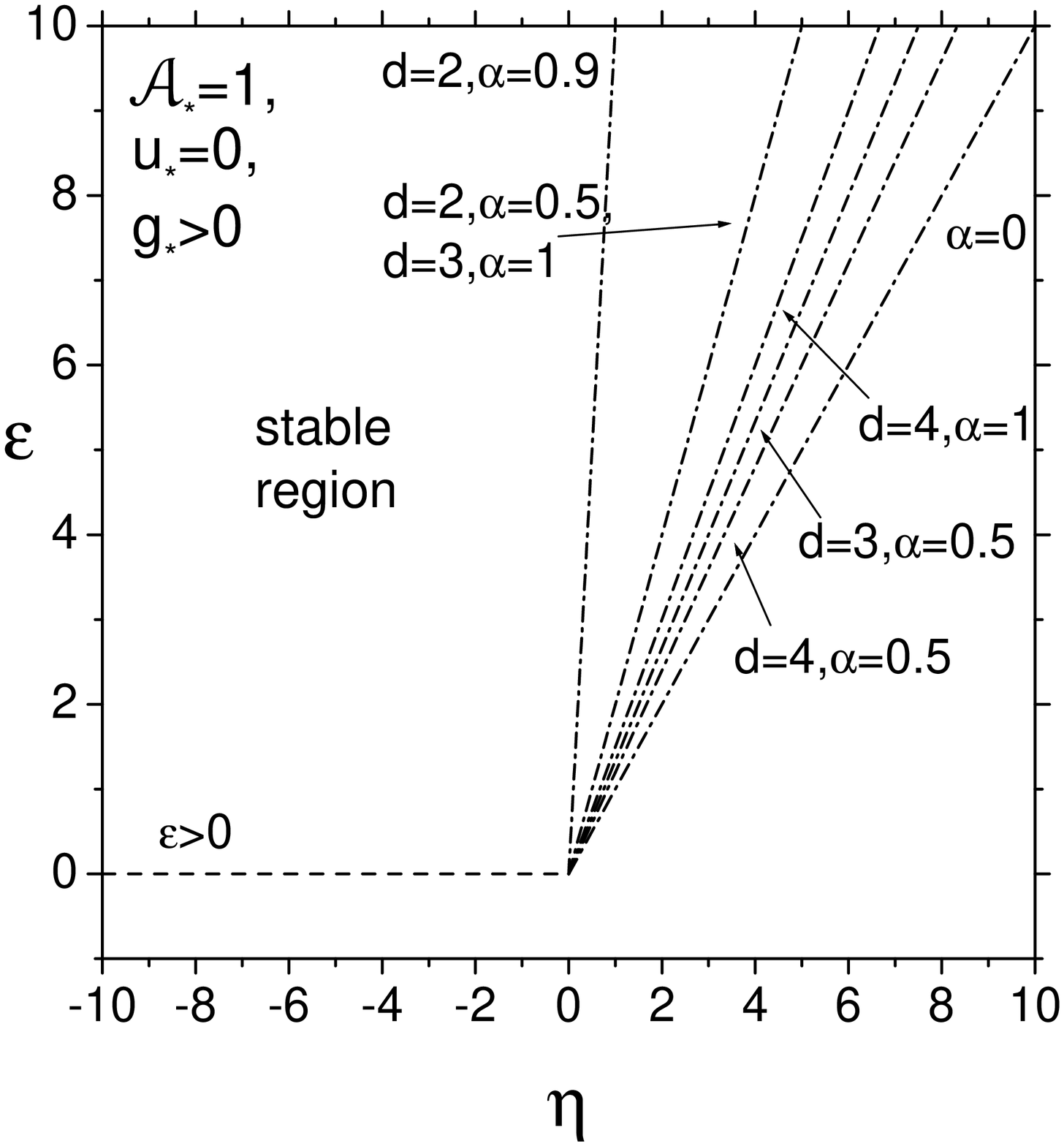,width=5cm}\hskip .5
cm \epsfig{file=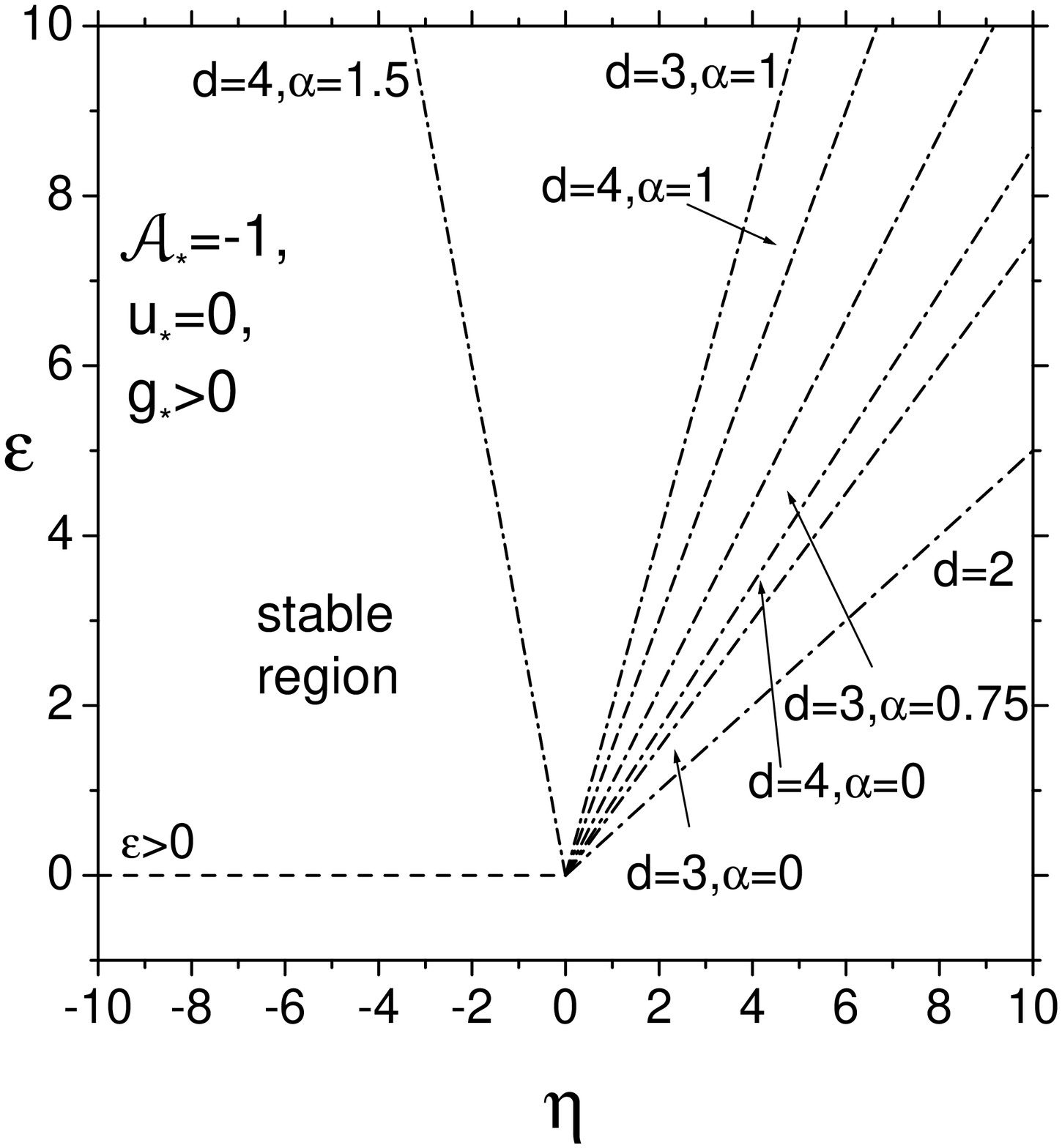,width=5cm}\hskip .5 cm
\epsfig{file=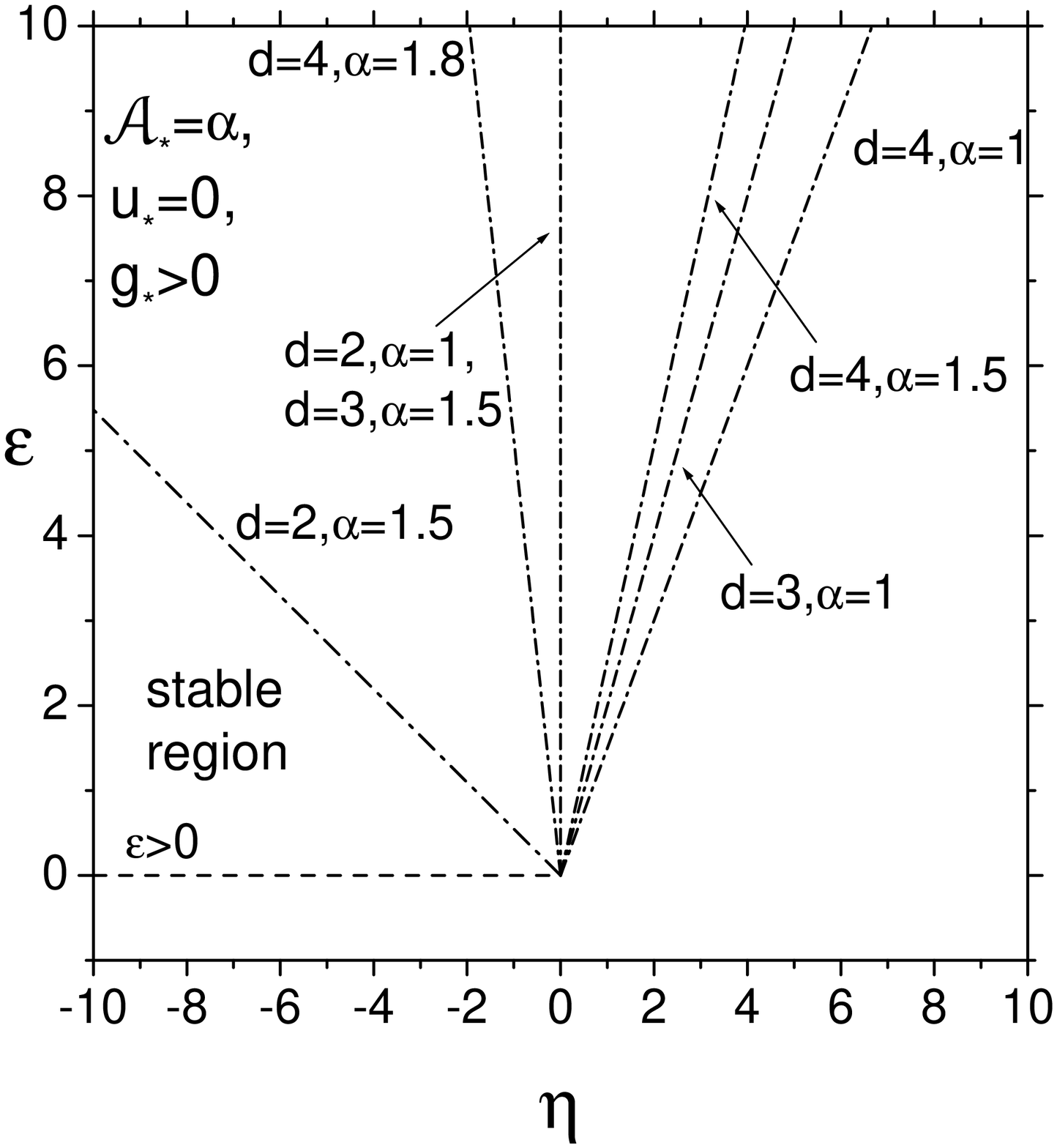,width=5cm}\hskip .5 cm}} \vspace{0.5cm}
\caption{Regions of infrared stability of the three fixed points
(\protect\ref{Q+}), (\protect\ref{Q-}), and (\protect\ref{Qa})
corresponding to quenched disorder for a representative set of
values of the space dimension $d$ and the parameter $\alpha$ of
the relative strength of the longitudinal part in the correlation
function of the velocity field. The region of stability for any
indicated values of $d$ and $\alpha$ lies between the dashed ray
($\varepsilon=0$, $\eta\le 0$) and the correspondingly marked
dash-dotted ray in the upper half of the $\eta$, $\varepsilon$
plane. Note different scales on the coordinate axes.} \label{Fig1}
\end{figure}

The admissibility regions in the $\eps$--$\eta$ plane are shown in
Figs. \ref{Fig1} for a number of values of the parameters $\alpha$ and $d$.
In the one-loop approximation, their boundaries are always given by
straight rays starting at the origin $\eps$--$\eta$. Some of them,
however, can be affected by the higher-order corrections, so that
the gaps or overlaps of different regions can appear in the two-loop
approximation.

It is interesting to note that the scaling regimes that arise as solutions
of the RG equations for the general model, in several cases correspond to
interesting physical situations. In particular, the regimes governed by the
points $Q^{\pm,\, \alpha}$ correspond to the case of time-independent (or
frozen) velocity field, while $F^{\pm,\, \alpha}$ correspond to the
rapid-change limit of the general model (\ref{f}).

To avoid possible misunderstandings we emphasize that the limits
$u_{0}\to0$ or $g_{0}\to\infty$, $u_{0}\to\infty$ are not supposed to be
performed in the original correlation function (\ref{f}); the parameters
$g_{0}$, $u_{0}$ (and hence $g$, $u$) are fixed at some finite values.
The behavior specific to the models (\ref{RC1}), (\ref{RC2}) arises
asymptotically as a result of the solution of the RG equations, when the
RG flow approaches the corresponding fixed point.
This shows that in the regimes governed by the points $Q^{\pm,\, \alpha}$,
the temporal fluctuations of the velocity field are asymptotically
irrelevant in determining the inertial-range behavior of the passive
field, which is then completely determined by the equal-time velocity
statistics. In the regimes governed by $R^{\A}$,
spatial and temporal fluctuations are both relevant,
but the effective correlation time of the passive field becomes so
large under renormalization that the correlation time of the velocity
can be completely neglected. The inertial-range behavior of the passive
field is determined solely by the $\omega=0$ mode of the velocity field;
this is the case of the rapid-change model.
In particular, this means that the coordinates of the fixed points and
the anomalous exponents in such regimes must depend on the only exponent
$\zeta\equiv 2\eps-\eta$ or $\eps$ that survives in the limit in question,
and coincide with the corresponding dimensions obtained directly for the
models (\ref{RC1}) or (\ref{RC2}). This is indeed the case, as one can
see from the explicit expressions given above and in the next Section.

As regards the value of the amplitude factor $\A$ in Eq. (\ref{1}), it can
remain an arbitrary parameter (for $R^{\A}$), or can be attracted by one of
the fixed points $\A_{*}=\pm 1$, or $\alpha$ (for $Q$ or $F$);
see Eq. (\ref{A}). Again, these possibilities correspond to
physical situations interesting as such.
The case $\A_{*}=1$ corresponds to the behavior
characteristic of the magnetic model, where the pressure vanishes due
to the relation (\ref{Pressure}) and the equation (\ref{1}) becomes local.
Therefore, the infrared behavior of the nonlocal model can be described
by the same fixed point (or universality class) as that of the local
(magnetic) model; the nonlocal pressure term does not affect the asymptotic
properties of the passive field. The general model indeed becomes
a ``turbulence  without pressure.'' The case $\A_{*}=-1$ corresponds to the
linearized NS equation with a given statistics of the background field
(we recall, however, that this value of $\A_{*}$ can be affected by the
higher-order corrections in $\eps$ and $\eta$). Finally, the case
$\A_{*}=0$, $\alpha=0$ corresponds the model, which (in its rapid-change
variant) was introduced independently in a number of studies as an example
of a linear system with pressure \cite{AP,vector},
a model with nontrivial mixing of composite operators \cite{LA,vector}
or a model which, with a proper choice of the forcing, can reproduce the
anomalous exponents of the NS velocity field  \cite{BBT}.

It is worth noting that for $\alpha\ne0$, the model with $\A_{0}=0$ is not
renormalizable, as follows from the analysis given in Sec.~\ref{sec:RG}.
That is, the second nonlinear term $V_i^{(2)}$ will be generated by the
renormalization procedure, even if it was absent in the original equation
(\ref{1}). Of course, this fact does not mean that the model with $\A_{0}=0$
and $\alpha\ne0$ is inconsistent; it rather means that its IR behavior is
described by one of the fixed point with $\A_{*}\ne0$.

\section{Critical scaling. Critical dimensions of composite operators}
\label{sec:Scaling}

Consider for definiteness some equal-time two-point quantity $F(r)$ that
depends on a single distance parameter $r$, for example, the pair
correlation function of the primary fields $\theta$, $\theta'$ or some
composite operators. We assume that $F(r)$ is multiplicatively
renormalizable, i.e., $F=Z_{F}F^{R}$ with certain renormalization
constant $Z_{F}$. Then the function $F^{R}(r)$ satisfies the RG equation
of the form $[{\cal D}_{RG}+\gamma_{F}]\,F(r)= 0$ with the operator
${\cal D}_{RG}$ from (\ref{RG2}) and $\gamma_{F} \equiv \Dm \ln Z_{F}$,
cf. (\ref{RGF1}). The functions $F$ and $F^{R}$ are equally suitable for
studying the asymptotic behavior: the difference is in the normalization,
choice of parameters (bare or renormalized) and the form of the perturbation
theory (in $g_{0}$ or in $g$). The solution of the RG equation can be written
in terms of the invariant variables introduced in (\ref{RGI}). The analysis
shows that in the IR asymptotic region, defined by the inequality
$\Lambda r \gg 1$ with $\Lambda$ from (\ref{gg}) and any fixed $mr$
with $m=1/L$ from (\ref{2}), the invariant charges approach one of the
IR attractive fixed points (the choice of the appropriate fixed point
is discussed in the previous Section), and, as a result,
the function $F(r)$ takes on the self-similar form
\begin{equation}
F(r) \simeq  \kappa_{0}^{d_{F}^{\omega}}\, \Lambda^{d_{F}}
(\Lambda r)^{-\Delta_{F}}\, \xi(mr),
\label{RGR}
\end{equation}
where $d_{F}^{\omega}$ and $d_{F}$ are the frequency and total canonical
dimensions of $F$, respectively (see Sec.~\ref{sec:RG} and
Table~\ref{table1}), and $\xi$ is some function whose explicit form
is not determined by the RG equation itself. The critical dimension
$\Delta_{F}$ of the quantity $F$ is given by the expression
\begin{equation}
\Delta_{F} = d_{F}^{k}+ \Delta_{\omega} d_{F}^{\omega}+\gamma_{F}^{*}
= d_{F} - \gamma^{*}_{\kappa} d_{F}^{\omega}+\gamma_{F}^{*},
\label{32B}
\end{equation}
where $\gamma_{F}^{*}$ denotes the value of the anomalous dimension
$\gamma_{F}$ at the fixed point in question, and
$\Delta_{\omega}=2-\gamma^{*}_{\kappa}$ with $\gamma_{\kappa}$
from (\ref{RG2}) is the critical dimension of frequency.

Each nontrivial fixed point $Q^{\pm,\, \alpha}$, $F^{\pm,\, \alpha}$ and
$R^{\A}$ from Sec.~\ref{sec:Fixed} corresponds to the scaling representation
of the form (\ref{RGR}) with its own set of critical dimensions $\Delta_{F}$
for all quantities $F$ and $\Delta_{\omega}$. In general, these dimensions
are infinite series in $\eps$ and $\eta$. For the rapid-change regime (that
is, for the point $R^{\A}$) they depend on the only exponent $\zeta\equiv
2\eps-\eta$ from (\ref{RC1}), while for the regimes with quenched disorder
(that is, for the points $Q^{\pm,\,\alpha}$) they depend on the only exponent
$\eps$ that survives in the limit (\ref{RC2}). For the fixed point $R^{\A}$,
the equation $\beta_{x}=0$ determines the values of $\gamma^{*}_{\kappa}=
\zeta$ and $\Delta_{\omega}=2-\zeta$ exactly, that is, without corrections
of order $\zeta^{2}$ and higher. This follows from the explicit expressions
(\ref{betas}), (\ref{betax}) and the vanishing of the anomalous dimensions
$\gamma_{1,2}$; see the discussion below Eq.~(\ref{betax}). For the fixed
points $F^{\pm,\,\alpha}$ with $u_{*}\ne 0$ the equation $\beta_{u}=0$ leads
to the exact result $\gamma^{*}_{\kappa}=\eta$; see Eq. (\ref{betas}). For
the fixed point $Q^{\alpha}$ with ${\cal A}_{*}=\alpha=0$, the exact result
$\gamma^{*}_{\kappa}=\eps$ follows from the equation $\beta_{g}=0$ and
vanishing of $\gamma_{1}$; see the discussion in Sec.~\ref{sec:RG}. For the
other regimes, the first-order expressions for $\gamma^{*}_{\kappa}$ are
directly obtained by substituting the coordinates of the fixed points
(\ref{Q+}), (\ref{Q-}), (\ref{Qa}) into the one-loop expression
(\ref{gammanu}) for $\gamma_{\kappa}$. The results can be summarized as
follows:
\begin{equation}
\gamma^{*}_{\kappa} =
\cases{ \zeta\equiv 2\eps-\eta \quad {\rm (exact)} &  for $R^{\A}$, \cr
\eta \quad {\rm (exact)}   &  for $F^{\pm,\,\alpha}$, \cr
\eps\, \displaystyle{\frac{(d-1-\alpha)}{(d-1)}}  &  for $Q^{+}$, \cr
\eps\, \displaystyle{\frac{(d^{2}-d-3d\alpha+2\alpha+2)}{(d-1)(d-2\alpha)}}
&   for $Q^{-}$, \cr
\eps\, \displaystyle{\frac{(d^{2}-d+2)}{d(d-1)}}
& for $Q^{-}$ and $\alpha=0$, \cr
\eps\, \displaystyle{\frac{(d^{2}-d\alpha^{3}+d\alpha^{2}-\alpha^{3}+\alpha^{2}-\alpha-3)}
{(d^{2}-d\alpha^{3}+d\alpha^{2} +d\alpha -\alpha^{3}
+\alpha^{2}+\alpha-3)}}  &   for $Q^{\alpha}$, \cr
\eps \quad {\rm (exact)}  &  for $Q^{\alpha}$ and $\alpha=0$. \cr }
\label{DeltaOmega}
\end{equation}
Then the dimensions $\Delta_{F}$, e.g., of
the primary fields $\theta$, $\theta'$ are obtained from
Eq. (\ref{32B}) and the data from Table~\ref{table1},
\begin{equation}
\Delta_{\theta} = -1 + \gamma^{*}_{\kappa}/2,  \quad
\Delta_{\theta'} = d- \gamma^{*}_{\kappa}/2,
\label{DeltaT}
\end{equation}
and the dimensions of their correlation functions are given by simple
sums over the fields entering into the function.

In the following, the crucial role will be played by the critical
dimensions $\Delta[n,l]$ associated with the irreducible
tensor composite fields
(``local composite operators'' in the field theoretic terminology)
built solely of the fields $\theta$ at a single spacetime point
$x=\{t,{\bf x}\}$. They have the form
\begin{equation}
F[n,l]\equiv \theta_{i_{1}}(x)\cdots \theta_{i_{l}}(x)\,
\left(\theta_{i}(x)\theta_{i}(x)\right)^{p} + \dots,
\label{Fnl}
\end{equation}
where $l\le n$ is the number of the free vector indices and $n=l+2p$ is
the total number of the fields $\theta$ entering into the
operator; the vector indices and the argument $x$ of the symbol
$F[n,l]$ are omitted. The dots $\dots$ stand for the appropriate
subtractions involving the Kronecker delta symbols, which ensure
that the resulting expressions are traceless with respect to
contraction of any given pair of indices, for example,
$\theta_{i}\theta_{j} - \delta_{ij}\theta_{k}\theta_{k}/d$, $
\theta_{i}\theta_{j} \theta_{k} - (\delta _{ij}\theta_{k}  +
\delta _{ik}\theta_{j} + \delta _{jk}\theta_{i})\theta^2 /(d+2)$ and so
on. We also note that the numbers $n$ and $l$ are even or odd
simultaneously.

Owing to the coincidence of the arguments, additional UV divergences
arise in the correlation functions involving such operators. They are
eliminated by means of the additional renormalization procedure, which
gives rise to new (independent of $Z_{1,2,\kappa}$ from (\ref{Rac}))
renormalization constants. The analysis similar to that given in Refs.
\cite{RG3,RG4} for the scalar and in \cite{Lanotte2} for the magnetic
model shows that, in the case at hand, these constants can be calculated
in the model without forcing (the bare propagator
$\langle \theta_{i} \theta_{j}\rangle _0$ from Eq. (\ref{lines})
does not enter into the corresponding Feynman diagrams). Then the
operators (\ref{Fnl}) appear multiplicatively renormalizable:
$F[n,l]=Z[n,l] \,F^{R}[n,l]$.

We have calculated all the constants $Z[n,l]$ in the one-loop
approximation (first order in $g$) in the MS scheme for the
special case $\eta=0$ and arbitrary $\eps$, which is sufficient to
find the corresponding anomalous dimensions $\gamma[n,l]=\Dm\ln
Z[n,l]$; cf. the discussion above Eq. (\ref{Z}). We omit the
calculation (which is very similar to that performed in Refs.
\cite{RG3,RG4,Lanotte2} for the scalar and magnetic cases) and
give only the final result:
\begin{eqnarray}
Z[n,l] = 1 + \frac{g \bar S_{d}} {8\eps\,d(d+2)\,(u+1)}
\left[ {\cal A}^{2} Q_{1} + \alpha Q_{2} \right],
\label{Znl}
\end{eqnarray}
where
\begin{eqnarray}
Q_{1}&=& n(d+n)(d-1) - l(l+d-2)(d+1),
\nonumber \\
Q_{2}&=& n(dn+n-d)(d-1) - l(l+d-2).
\label{Qs}
\end{eqnarray}
For the anomalous dimension we thus obtain:
\begin{equation}
\gamma[n,l]= \frac{-g \bar S_{d}} {4d(d+2)\,(u+1)}
\left[ {\cal A}^{2} Q_{1} + \alpha Q_{2} \right].
\label{Gnl}
\end{equation}
According to Eq. (\ref{32B}), the critical dimensions of the operators
(\ref{Fnl}) are given by
\begin{equation}
\Delta[n,l]= n \Delta_{\theta} + \gamma^{*}[n,l]
\label{DnL}
\end{equation}
with $\Delta_{\theta}$ from (\ref{DeltaT}) and  $\gamma^{*}[n,l]$ is the
value of the anomalous dimension (\ref{Gnl}) at one of the nontrivial
fixed points (\ref{Q+}), (\ref{Q-}), (\ref{Qa}), (\ref{F+}), (\ref{F-}),
(\ref{Fa}), (\ref{RCP}). Note that owing to the renormalization,
the critical dimensions of the operators (\ref{Fnl}) differ from
the naive sum of the dimensions of the fields $\theta$ that constitute
the operator. The exception is provided by the case ${\cal A}_{*}=\alpha=0$,
when $\Delta[n,l]= n \Delta_{\theta}$ exactly (the proof is similar to
that given, e.g., in Ref. \cite{RG} for the scalar case).

The results for the anomalous dimensions $\gamma^{*}[n,l]$ can be
summarized as follows:
\[ \gamma^{*}[n,l] =
-(2\eps-\eta)\left[{\cal A}_{*}^{2}Q_{1}+\alpha Q_{2}\right]/2\times \]
\begin{equation}
\times\cases{ \displaystyle{\frac{1}{(d+2)(d-1+\alpha)}}
&  for $F^{+}$, \cr
\displaystyle{\frac{1}{(d-2)(d+1-\alpha)}}
&  for $F^{-}$, \cr
\displaystyle{\frac{1}{
(d^{2}-d\alpha^{3}+d\alpha^{2}+2d\alpha-\alpha^{3}+\alpha^{2}+3\alpha-3)
}} &  for $F^{\alpha}$, \cr
\displaystyle{\frac{1}{(d^{2}-3)}} &  for $F^{\alpha}$
and $\alpha=0$ \cr}
\label{Anomalous1}
\end{equation}
for the regimes with finite correlation time,
\[ \gamma^{*}[n,l] = -\eps\,
\left[{\cal A}_{*}^{2}Q_{1}+\alpha Q_{2}\right]/2\times \]
\begin{equation}
\times\cases{ \displaystyle{\frac{1}{(d+2)(d-1)}} &  for $Q^{+}$, \cr
\displaystyle{\frac{1}{(d-1)(d-2\alpha)}} &  for $Q^{-}$, \cr
\displaystyle{\frac{1} {(
d^{2}-d\alpha^{3}+d\alpha^{2}+d\alpha-\alpha^{3}+\alpha^{2}+\alpha-3)}}
&  for $Q^{\alpha}$, \cr
\displaystyle{\frac{1} {(d^{2}-3)}}
&  for $Q^{\alpha}$ and $\alpha=0$ \cr}
\label{Anomalous2}
\end{equation}
for the regimes with quenched disorder and
\[ \gamma^{*}[n,l] =
-(2\eps-\eta)\left[{\cal A}^{2}Q_{1}+\alpha Q_{2}\right]/2\times \]
\begin{equation}
\times\cases{
\displaystyle{\frac{1}{(d^{2}+{\cal A}^{2}-\alpha {\cal A}^{2}
-\alpha d {\cal A}^{2}+ \alpha d {\cal A} + d {\cal A} + d\alpha
+3\alpha -3)}} &  for $R^{\A}$, \cr
\displaystyle{\frac{1}{(d^{2}+{\cal A}^{2}+ d {\cal A}-3)}}
&  for $R^{\A}$ and $\alpha=0$ \cr}
\label{Anomalous3}
\end{equation}
for the regimes with zero correlation time. We recall that ${\cal A}_{*}$
takes on the values $1$, $-1$ and $\alpha$ for the fixed points $F$ and $Q$
labelled by the superscripts $+$, $-$ and $\alpha$, respectively, while
for the rapid-change regime $\A_*={\cal A}$ remains an arbitrary parameter.
We also note that the dimensions (\ref{Anomalous2}) depend on the only
exponent $\eps$ that survives in the limit (\ref{RC2}), while the
dimensions (\ref{Anomalous3}) depend on the only exponent $\zeta=2\eps-\eta$
that survives in the limit (\ref{RC1}). The exponents (\ref{Anomalous1})
also depend only on $\zeta$, which seems to be an artifact of the
first-order approximation. We also note that the exponents (\ref{Anomalous3})
were derived earlier directly for the rapid-change model (\ref{RC1}) for
some special cases: ${\cal A}=1$ and $\alpha=0$ in \cite{Lanotte2},
${\cal A}=1$ and arbitrary $\alpha\ge0$ in \cite{alpha}
and arbitrary ${\cal A}$ and $\alpha=0$ in \cite{amodel}.

From Eqs. (\ref{DeltaOmega}), (\ref{DeltaT}) and
(\ref{DnL})--(\ref{Anomalous3}) it follows that
\begin{equation}
\Delta[n,l]=-n+O(\varepsilon).
\label{hier1}
\end{equation}
Thus for all nontrivial fixed points, at least for small values of
$\varepsilon\sim \eta$, one has $\Delta[n,l]<\Delta[k,j]$ if $n>k$
regardless of the relation between $l$ and $j$. For a fixed value
of $n$ one has
\begin{equation}
\Delta[n,l]>\Delta[nj]\quad  {\rm and} \quad \gamma[n,l]>\gamma[nj]
\quad {\rm if} \ l>j,
\label{hier2}
\end{equation}
as one can easily see from the original expression (\ref{Gnl}) for
the anomalous dimension $\gamma[n,l]$, properties of the
polynomials (\ref{Qs}) and the fact that the combination $g/(u+1)=
x/(w+1)$ that enters into Eq. (\ref{Gnl}) is positive definite for
all nontrivial fixed points. [We recall that the parameters $x$,
$w$ were introduced above Eq. (\ref{betax}) for the proper
description of the rapid-change regimes.] Thus for fixed $n$, the
dimension $\Delta[n,l]$ decreases monotonically with $l$ and
reaches its minimum for the minimal possible value of $l$, that
is, $l=0$ if $n$ is even and $l=1$ if $n$ is odd.

The hierarchy relations (\ref{hier1}), (\ref{hier2}) will be
important in the discussion of the inertial-range behavior of
various correlation functions, in particular, in the issue of the
large-scale anisotropy persistence; see Sec.~\ref{sec:OPE}.
Similar inequalities were established earlier for the case of the
passive scalar field, advected by the velocity ensemble (\ref{f}),
(\ref{Fin}) in Refs. \cite{RG3,RG4}, and for the magnetic field
advected by the Kraichnan ensemble (\ref{RC1}) in
\cite{Lanotte,Lanotte2}.

It also follows from (\ref{hier1}) that the critical dimensions
$\Delta[n,l]$ are {\it negative} and that the spectrum of their
dimensions is not bounded from below; these properties, typical of
the models of turbulence, will also be important in the following.

If the random force $f$ is introduced in Eq. (\ref{1}), none of
the formulas (\ref{Znl})--(\ref{Anomalous3}) change, because the
bare correlator $\langle\theta\theta\rangle_0$ from Eq.
(\ref{lines}), which becomes nonzero, does not enter the relevant
diagrams. However, the operators (\ref{Fnl}) are no longer
renormalized multiplicatively: the operator $F[k,j]$ can admix in
renormalization to the operator $F[n,l]$ if, and only if, $k<l$. In
general, $j\ne l$, but if the correlator (\ref{2}) is isotropic,
only operators with $j=l$ can mix in renormalization. The
admixture of operators with $k>l$ is impossible due to the absence
of appropriate diagrams; this is a consequence of the linearity of
the original equation (\ref{1}) in $\theta$ and $f$. The admixture
of operators with $k=n$ and $j\ne l$ is also impossible, because
the corresponding diagrams do not involve the correlator
$\langle\theta\theta\rangle_0$ and therefore do not ``feel'' the
violation of the rotational symmetry caused by the function (\ref{2}).

As a result of the mixing, the operator $F[n,l]$ becomes a finite
sum of contributions with definite critical dimensions
$\Delta[n,l]$ and $\Delta[k,j]$ with $k<n$ and, in general, all
possible values of $j$ allowed for a given $k$. However, due to
the relation (\ref{hier1}), the leading term is still given by the
original contribution with the dimension $\Delta[n,l]$, while the
new contributions with $k<n$ give only corrections that vanish in
the IR range $\Lambda r \gg 1$ in expressions like (\ref{RGR}). In
what follows, we shall be interested only in the leading terms of
and thus we can ignore the mixing and treat the operator $F[n,l]$
as if it has the definite critical dimension $\Delta[n,l]$.

\section{Operator product expansion and the anomalous scaling}
\label{sec:OPE}

The representation (\ref{RGR}) for any scaling function $\xi(mr)$
describes the behavior of the correlation function $F(r)$ for
$\Lambda r \gg1$ and any fixed value of $mr$. The inertial range
corresponds to the additional condition that $mr\ll 1$. The form
of the function $\xi(mr)$ is not determined by the RG equations
themselves; in the theory of critical phenomena, its behavior for
$mr\to0$ is studied using the well-known Wilson operator product
expansion (OPE); see, e.g., Ref.~\cite{Zinn}. This technique is
also applicable in the theory of turbulence; see, e.g.,
Ref.~\cite{JETP,UFN,turbo}.

According to the OPE, the equal-time product $F_{1}(x)F_{2}(x')$
of two renormalized composite operators at ${\bf x}\equiv ({\bf x}
+ {\bf x'} )/2 = {\const}$ and ${\bf r}\equiv {\bf x} - {\bf
x'}\to 0$ can be represented in the form
\begin{equation}
F_{1}(x)F_{2}(x')=\sum_{F} C_{F} ({\bf r}) F(t,{\bf x}),
\label{OPE}
\end{equation}
where the functions $C_{F}$  are the Wilson coefficients regular
in $m^{2}$ and $F$ are, in general, all possible renormalized local
composite operators allowed by symmetry; more precisely, the
operators entering into the OPE are those which appear in the
corresponding Taylor expansions, and also all possible operators
that admix to them in renormalization. If these operators have
additional vector indices, they are contracted with the
corresponding indices of the coefficients $C_{F}$.

Without loss of generality it can be assumed that the expansion in
Eq. (\ref{OPE}) is made in the operators with
definite critical dimensions $\Delta_{F}$. The renormalized
correlation function $\langle F_{1}(x)F_{2}(x') \rangle$
is obtained by averaging Eq. (\ref{OPE}) with the weight
$\exp S_{R}$ with $S_{R}$ from Eq. (\ref{Rac}), the quantities
$\langle F \rangle$
appear on the right hand side. Their asymptotic behavior
for $m\to0$ is found from the corresponding RG equations and
has the form $\langle F \rangle \propto  m^{\Delta_{F}}$.

From the operator product expansion (\ref{OPE}) we therefore
find the following expression for the scaling function
$\xi(mr)$ in the representation (\ref{RGR}) for the
correlation function $\langle F_{1}(x)F_{2}(x') \rangle$:
\begin{equation}
\xi(mr)=\sum_{F}A_{F}\,(mr)^{\Delta_{F}},
\label{OR}
\end{equation}
where the coefficients $A_{F}(mr)$ are regular in $(mr)^{2}$.

The quantities of interest are, in particular, the equal-time pair
correlation functions of the composite operators (\ref{Fnl}). For
these, the representation (\ref{RGR}) is valid with the dimensions
$d_{F}^{\omega}=-(n+k)/2$, $d_{F}=-(n+k)$ and $\Delta_{F}=
\Delta[n,l]+\Delta[k,j]$ with $\Delta[n,l]$ from (\ref{DnL}):
\begin{equation}
\langle  F[n,l](t,{\bf x})F[k,j](t,{\bf x}') \rangle =
(\kappa_0)^{-(n+k)/2} \Lambda^{-(n+k)} (\Lambda
r)^{-\Delta[n,l]-\Delta[k,j]}\, \xi(mr)
 \label{struc}
\end{equation}
where $\Lambda r \gg 1$ and $\xi(mr)$ is the corresponding scaling
function.

As already said above, the operators entering into the OPE are
those which appear in the corresponding Taylor expansions, and
also all possible operators that admix to them in renormalization.
The leading term of the Taylor expansion for the function
(\ref{struc}) is given by the $n$-th rank tensor $F[n+k,l+j]$ from
Eq. (\ref{Fnl}). Its decomposition in irreducible tensors gives
rise to the operators $F[n+k,p]$ with all possible values of $p\le
l+j$; the admixture of junior operators (see the end of Sec.
\ref{sec:Scaling}) gives rise to all the monomials $F[s,p]$  with
$s<n+k$ and all possible $p$ allowed for a given $s$. Hence, the
asymptotic expression for the structure function $\xi(mr)$ for
$mr\ll 1$ has the form
\begin{equation}
\xi(mr) = \sum_{s=0}^{n+k} \sum_{p=p_{s}}^{s} \Bigl[A_{sp}\,
(mr)^{\Delta[s,p]}+\dots\Bigr],
\label{struc2}
\end{equation}
with the dimensions $\Delta[k,p]$ from Eq. (\ref{DnL}). Here and
below $p_{s}$ denotes the minimal possible value of $p$ for given
$s$, i.e., $p_{s}=0$ for $k$ even and $p_{s}=1$ for $k$ odd;
$A_{sp}$ are some numerical coefficients dependent on the
parameters like $\eps$, $d$ and so on. The dots in Eq.
(\ref{struc2}) stand for the contributions which arise from the
composite operators that, in addition to the field $\theta$,
involve the other fields $\theta'$, ${\bf v}$ and/or derivatives
$\partial$, $\partial_t$.

The leading term of the expression (\ref{struc2}) for $mr\ll 1$ is
determined, obviously, by the minimal possible dimension
$\Delta_{F}$ that appear on its right-hand side, provided this
minimal dimension exists. In our model, there are infinitely many
operators with negative critical dimensions, and the spectrum of
their dimensions is not bounded from below. (It is possible to
show on general grounds that, if a model involves one negative
dimension, it necessarily involves infinitely many negative
dimensions with unbounded spectrum.) If all these operators
appeared on the right-hand side of the representation
(\ref{struc2}), we would have to sum up their contributions in
order to find the asymptotic behavior at $mr\to 0$. This problem
is indeed encountered for the stochastic NS equation \cite{JETP},
and is discussed in Refs. \cite{turbo,UFN} in detail.

In our model, however, there is no such problem, at least for
small $\varepsilon$. The contributions of the operators $F[s,p]$
with $s>n+k$ (which would be more important) do not appear in Eq.
(\ref{struc2}), because they are absent in the Taylor expansion of
the correlator (\ref{struc}) and do not admix in renormalization
to the terms of the Taylor expansion; see Sec. \ref{sec:Scaling}.
As already noted there, this is a manifestation of the linearity
of the original equation (\ref{1}) in $\theta$ and $f$. What is
more, one can show that for any operator $F$ that appear in the
OPE (and not only the operators (\ref{Fnl}) built solely of the
fields $\theta$) the number of the fields $\theta$ cannot exceed
the total number of the fields $\theta$ on the left-hand side;
therefore their dimensions cannot appear in (\ref{struc2}). It
then follows that the leading term in (\ref{struc2}) is determined
by an operator built solely of the fields $\theta$ and containing
the maximal possible number of the fields, that is, $n+k$. The
operators containing less than $n+k$ fields $\theta$ give only
corrections, as follows from the hierarchy relation (\ref{hier1}).
The operators involving the fields $\theta'$, ${\bf v}$ and/or
derivatives also give only corrections, because the canonical
dimensions of these additional factors are positive (see
Table~\ref{table1}) and thus increase the total canonical
dimension of the operator in comparison with the corresponding
operator built solely of the fields $\theta$. Furthermore, from
the hierarchy relation (\ref{hier2}) it follows that, for the
fixed number of the fields $\theta$, the minimum of the dimension
is achieved for the minimal number of vector indices, that is, for
the scalar operator $F[n+k,0]$ if the sum $n+k$ is even and the
vector operator $F[n+k,1]$ if the sum $n+k$ is odd.

We therefore conclude that the leading term of the small-$mr$
behavior of the scaling function (\ref{struc2}) has the form $\xi
\sim (mr)^{\Delta[n+k,l_{n+k}]}$. Substituting this expression
into Eq. (\ref{struc}) gives the desired leading term of the
correlation function of two operators (\ref{Fnl}) in the
inertial-range ($\Lambda r \gg 1$, $mr\ll 1$):
\begin{equation}
\langle  F[n,l](t,{\bf x})F[k,j](t,{\bf x}') \rangle \sim
(\kappa_0)^{-(n+k)/2} \Lambda^{-(n+k)} (\Lambda
r)^{-\Delta[n,l]-\Delta[k,j]}\, (mr)^{\Delta[n+k,l_{n+k}]}
\label{struc4}
\end{equation}
with the dimensions $\Delta[n,l]$ from (\ref{DnL}).

\section{Anomalous scaling in anisotropic sectors. Hierarchy
of anisotropic contributions} \label{sec:Hier}

Without loss of generality, it can always be assumed that the expansion
(\ref{OPE}) is made in irreducible traceless tensor composite operators.
Then averaging Eq. (\ref{OPE}) with the weight $\exp S_{R}$ automatically
produces the decomposition of the correlation function in irreducible
representations of the rotation group $SO(d)$, similar to that employed e.g.
in Refs. \cite{Arad1,Arad2,Arad3,Arad4,BDAT,BCTT} for description of the NS
turbulence. This becomes especially clear if the left-hand side of
Eq. (\ref{OPE}) involves only scalar quantities and the anisotropy,
introduced by the correlator (\ref{2}), is uniaxial, that is, specified
by a single constant unit vector ${\bf n}$. Then the mean value
$\langle F\rangle$ of a $l$-th rank tensor operator $F$ on the right-hand
side of (\ref{OPE}) is an irreducible traceless $l$-th rank tensor built
only of the vector ${\bf n}$ and the Kronecker delta symbols. Its vector
indices are contracted with the indices of the corresponding Wilson
coefficient $C_{F}({\bf r})$, which gives rise to the Legendre (or
Gegenbauer for arbitrary $d$) polynomial of order $l$. In general,
decomposition in (hyper)spherical harmonics (see e.g. \cite{harm} and
references therein) or its analogs for tensor quantities (see e.g.
\cite{OR} and the references) will be encountered.

The rank $l$ of the operator can be viewed as the measure of
anisotropy of the corresponding contribution in expansion (\ref{struc2}).
If the forcing is isotropic, that is, the function $C({\bf r})$ in the
correlator (\ref{2}) depends only on $r=|{\bf r}|$, only scalar operators
with $l=0$ have nonvanishing mean values, and only their dimensions
appear on the right-hand side of Eq. (\ref{struc2}). In general,
tensor operators with $l\ne0$ also contribute to (\ref{struc2}).
Owing to the relations (\ref{hier2}), the leading term of the asymptotic
behavior at $mr\to0$ is still given by the scalar operator with $l=0$
(it has the minimal dimension among the operators with a fixed number of
the fields). We thus conclude that the leading term is given by the
same expression (\ref{struc4}) for both the isotropic and anisotropic
forcing, while anisotropic contributions with $l>0$ give only subleading
terms (corrections). What is more, relations (\ref{hier2}) show that these
contributions reveal a kind of {\it hierarchy} related to the degree of
anisotropy: the higher is the rank of the operator, the less important is
its contribution to the inertial-range behavior.

For the first time, the hierarchy relations for anisotropic contributions
were derived in Ref. \cite{Lanotte} for the magnetic field, passively
advected by Kraichnan's velocity ensemble (\ref{RC1}), and in Ref. \cite{RG3}
for the scalar field, advected by the Gaussian velocity field specified by
the correlator (\ref{Fin}) (in both cases with general $d$ and $\alpha=0$).
In the first of these papers, anomalous exponents were found exactly for the
pair correlation function, while in the second the exponents were derived
only in the one-loop approximation, but for all the higher-order correlation
functions. Later these results were reproduced in Ref. \cite{AB} for the
magnetic field (only for $d=3$, but also including helical contributions)
and in \cite{ST} for the scalar field advected by Kraichnan's ensemble.
Generalization to the higher-order correlation functions of the magnetic
field was given in \cite{Lanotte2} (for $\alpha=0$), while generalizations
to the case of general $\alpha$ were obtained in \cite{alpha} (magnetic
field and Kraichnan's ensemble) and in \cite{RG4} (scalar field and the
ensemble (\ref{Fin})). Generalization to the general vector model (\ref{1})
and Kraichnan's ensemble was given in \cite{amodel}.

So far, analytical results of such kind have been obtained only for passive
fields, advected by the synthetic Gaussian velocity ensembles. However,
numerical simulations and real experiments show that the picture outlined
above appears rather general, being observed by the passive scalar field
advected by the two-dimensional NS field in the inverse energy cascade
\cite{fronts} and by the NS velocity field itself
\cite{Arad3,Arad4,BDAT,BCTT}.
These observations justify and make more precise old phenomenological ideas
about the isotropization of the inertial-range turbulence in the presence
of a large-scale anisotropy. Nevertheless, the anisotropy survives
in the inertial range and reveals itself in {\it odd} correlation functions,
in disagreement with what was expected on the basis of the cascade ideas.
We shall return to this important issue in the Conclusion, and now
let us briefly discuss the influence of compressibility on the hierarchy
of the anomalous exponents.

Effects of the compressibility on the anomalous scaling in anisotropic
sectors were studied earlier for the scalar \cite{Komp2} and magnetic
\cite{alpha} fields advected by Kraichnan's ensemble (\ref{RC1}) and
for the scalar advected by the velocity ensemble (\ref{Fin}) with finite
correlation time \cite{RG4}; see also Ref. \cite{alpha2} for a summary.
In all those cases the conclusion was the same: the hierarchy expressed
by the relation (\ref{hier2}), which can be rewritten as
\begin{equation}
\partial \Delta[n,p] / \partial p>0,
\label{DER1}
\end{equation}
remains valid for all values of the compressibility parameter
$0\le \alpha <+\infty$, but it always becomes less pronounced
as $\alpha$ grows,
\begin{equation}
\partial^{2} \Delta[n,p] / \partial p \, \partial\alpha<0.
\label{DER2}
\end{equation}
This means, in particular, that the anisotropic corrections in Eq.
(\ref{struc2}) become closer to each other and to the leading term
as $\alpha$ grows. Thus the compressibility enhances the penetration of
the large-scale anisotropy into the inertial range. This penetration is
even more manifest for the odd-order ratios of the correlation functions:
the skewness factor grows for $mr\to0$, provided $\alpha$ is large enough,
while the growth of the hyperskewness factor and other higher-order ratios
becomes much faster than for the incompressible case; see the discussion
in \cite{Komp2,RG4,alpha,alpha2}.

No such definite conclusions can be drawn for the general vector model.
The straightforward analysis of the explicit expressions
(\ref{Anomalous1})--(\ref{Anomalous3}) shows that the
derivative $\partial^{2} \Delta[n,p] / \partial p \partial\alpha$
is negative and, therefore, the behavior described above takes place
only in two regimes described by the fixed points $F^{+}$ (always)
and $R^{\A}$ (only if the relation
\begin{equation}
-3+\A d +d^{2}-\A^3 d (d+1)+\A^4 (d+1)^{2}-\A^2 (d^{2}+4d+2) >0,
\label{XPEH}
\end{equation}
which is independent on $\alpha$, is satisfied). For all the other cases,
one finds $\partial^{2} \Delta[n,p] / \partial p \partial\alpha>0$ and
the behavior is opposite: compressibility suppresses
the penetration of the large-scale anisotropy into the inertial range,
anisotropic contributions become further from one another and from the
isotropic term.

It is tempting to attribute this ``inverse behavior'' to the combined
influence of the compressibility and pressure. Indeed, the ``normal''
(scalar-like) behavior takes place for the magnetic regime $F^{+}$,
for which $\A_*=\A_{0}=1$ and the pressure term (\ref{Pressure}) vanishes.
For the rapid-change regime $R^{\A}$ and reasonable values of $d$, the
relation (\ref{XPEH}) is satisfied only in the restricted area around the
point $\A_*=1$ (including $\A_*=1$, in agreement with the analysis of
\cite{alpha}), where the pressure effects are relatively small. However,
the magnetic regime in a frozen velocity field, $Q^{+}$, demonstrates the
inverse behavior.

The dependence on the parameter $\A$, that controls the pressure effects, is
essentially different from the dependence on $\alpha$: for the regimes with
nonzero (finite or infinite) correlation time, the value of the corresponding
invariant variable can take only three discrete values $\A_{*}=-1$, 1,
$\alpha$. The value of $\A_{*}$ which is realized for a given regime depends
on $\alpha$ but not on $\A$; see Eq. (\ref{A}). The case $\A_{*}=1$
corresponds to the magnetic (pressureless) equations; then the general model
indeed becomes a ``turbulence  without pressure,'' despite the presence of
the nonlocal pressure term in the original stochastic equation. For the
zero-correlated regime $R^{\A}$, the anomalous exponents retain a continuous
dependence on $\A$; for the incompressible case ($\alpha=0$) it was discussed
in \cite{amodel}. The value of $\A=1$, where the pressure effects disappear,
is not distinguished at all; the derivative $\partial^{2} \Delta[n,p] /
\partial p \partial\A$  at $\A=1$ is positive for almost all parameters
(namely, for $d> -1.3 \alpha+1.5$, the approximate relation obtained
numerically), but is negative definite e.g. for $\A=0$.

\section{Conclusion} \label{sec:Conc}

We have studied a model of a divergence-free (transverse) vector quantity
passively advected by a random Gaussian velocity field with finite (and not
small) correlation time. The model is described by an advection-diffusion
equation with a random large-scale stirring force, nonlocal pressure term
and the most general form of the inertial nonlinearity.
The correlation function of the advecting field mimics
some properties of the real inertial-range turbulence: the energy spectrum
has the form $E(k)\propto k^{1-2\eps}$, while the correlation time scales as
$k^{-2+\eta}$. An advantage of the model is the possibility to control the
pressure contribution and thus study its effects on the inertial-range
behavior. Another reason to study the general case is the possibility to
describe in a uniform way several special cases interesting on their own:
the kinematic magnetic model, linearized NS equation and the special
model without the stretching term, which possesses additional symmetry
and has a close formal resemblance with the nonlinear NS case.

We have shown that the system exhibits various types of inertial-range
asymptotic behavior, characterized by nontrivial anomalous exponents;
the latter are analytically calculated to first order in $\eps\sim\eta$,
including the anisotropic sectors.

The key points of our analysis are the existence of a field theoretic
formulation of the original stochastic problem (Sec.~\ref{sec:FT}),
multiplicative renormalizability of the corresponding field theory
(Sec.~\ref{sec:RG}), existence of nontrivial IR-attractive fixed points
of the corresponding RG equations in the physical region of the parameters
(Sec.~\ref{sec:Fixed}) and the possibility to identify the anomalous
exponents with the critical dimensions of certain composite operators
(Secs.~\ref{sec:Scaling} and~\ref{sec:OPE}). This allows one to construct
a systematic perturbation expansion for the exponents; the practical
calculations have been performed to the first nontrivial order in
$\eps\sim\eta$ (one-loop approximation).

Existence of explicit one-loop expression allows one to discuss the
stability scaling regimes and the universality of the corresponding
exponents, that is, their (in)dependence on the pressure, anisotropy,
compressibility, forcing and so on, or, more technically, on the
exponents $\eps$, $\eta$ and the amplitudes $\A_{0}$, $g_{0}$, $u_{0}$
and $\alpha$ in the stochastic equation (\ref{1})
and the correlator (\ref{Fin}) of the advecting velocity. Although
the behavior of the vector model is much richer than that of its
scalar counterpart, the general picture appears essentially the same:
the exponents are universal in the sense that they depend on the
exponents $\eps$ and $\eta$, but do not depend on the amplitudes in
(\ref{Fin}) and the forcing (\ref{2}); the exponents related to
anisotropic contributions show a hierarchy related to the degree of
anisotropy (more anisotropic contributions are less important);
this hierarchy holds for all scaling regimes, regardless of the values
of the compressibility parameter $\alpha$ from (\ref{f}) and
the pressure parameter $\A$ from (\ref{1}). Consider these points
in more detail.

{\bf Scaling regimes and universality classes.} Infrared asymptotic behavior
of our model is completely described by seven different scaling regimes, or
universality classes, each corresponding to a set of anomalous exponents.
For the given set of the parameters $\eps$, $\eta$ and $\alpha$, only one of
these regimes can be realized, irrespective of the values of the amplitudes
$\A_{0}$, $g_{0}$, $u_{0}$. Three regimes correspond to finite correlation
time of the advecting velocity field, and three regimes correspond to
infinite correlation time (or time-independent velocity). Each of these two
sets involve the magnetic (pressureless) case, linearized NS equation and the
model with $\alpha$-dependent effective amplitude in front of the stretching
term; for $\alpha=0$ the latter gives the special model with no stretching
term, whose rapid-change version was studied e.g. in \cite{vector,AP,BBT}.
The scaling exponents for these regimes depend on the exponents $\eps$,
$\eta$ and the amplitude $\alpha$, but they are independent of the values
of the amplitudes $\A_{0}$, $g_{0}$, $u_{0}$. The remaining seventh regime
corresponds to the rapid-change velocity (zero correlation time), the
corresponding exponents depend also on $\A_{0}$.

To avoid possible confusion we stress that the behavior specific to the
aforementioned classes, e.g. magnetic model with infinite correlation time,
arises automatically when the RG flow approaches the fixed point which is IR
attractive for the given choice of parameters $\eps$, $\eta$, $\alpha$; the
frozen limit (\ref{RC2}) or the substitution $\A_{0}=1$ are not performed in the
original model and the parameters $\A_{0}$, $g_{0}$, $u_{0}$ are fixed at
such finite values. In particular this means that the anomalous exponents
in those regimes are independent of the correlation time (more precisely,
the ratio of the correlation time of the velocity field and the turnover
time for the scalar field, measured by the parameter $u_{0}$; see the
discussion in \cite{RG3}). In this sense, one can speak about the
universality of the anomalous exponents in our model.

{\bf Independence of the forcing. Zero-mode picture.}
As we have seen, the critical dimensions of all composite operators
(\ref{DnL}), and therefore the corresponding anomalous exponents (including
anisotropic sectors), are independent of the forcing, specified by the
correlator (\ref{2}). In particular, this means that they remain unchanged,
when the stirring force in Eq. (\ref{1}) is replaced by the imposed mean
constant field, like in Refs. \cite{Lanotte2,Lanotte}. The role of the
forcing is to maintain the steady state of the system and thus to provide
nonzero {\it amplitudes} for the power-like terms with those universal
exponents.

This behavior is already well known for the passive scalar fields
\cite{RG3,RG4} advected by the velocity (\ref{Fin}) or vector fields,
advected by the zero-correlated velocity \cite{Lanotte2,Lanotte}.

In the language of the RG (which is equally applicable to the case of a
zero or finite correlation time) this is explained as follows: the stirring
force or the mean field do not enter into the diagrams that determine the
renormalization of the operators (\ref{DnL}), so that their dimensions are
independent of the forcing. Similar diagrams determine the contributions
of those operators into the operator-product expansions (\ref{OPE}), which
are nontrivial even for the unforced model. The difference is that for the
unforced model, mean values of the operators vanish, and they give no
contribution to the right-hand sides of representations like (\ref{struc2}).
For the isotropic correlator (\ref{2}), scalar operators acquire nonzero
mean values and contribute to the right-hand side of (\ref{struc2}), while
for the anisotropic correlator or the imposed mean field, the mean values
of irreducible tensor operators also become nonzero and their
contributions are ``activated'' in representations (\ref{struc2}).

For the case of zero correlation time, when the equal-time correlations
functions satisfy exact closed differential equations, the above
picture it is easily understood in the language of the zero-mode approach
\cite{FGV}: forcing terms do not affect the corresponding differential
operators; thus the anomalous exponents, determined by the zero modes
(solutions of homogeneous unforced equations) also are independent of
the forcing. On the contrary, the {\it amplitudes} are determined by the
matching of the inertial-range zero-mode solution with the forced large-scale
solutions, which is only possible in the presence of the forcing terms.

The exact resemblance in the behavior of the rapid-change models and the
finite-correlated cases suggests that for the latter, the concept of zero
modes (and thus of statistical conservation laws) is also applicable,
although the corresponding equations are not differential and involve
infinite diagrammatic series.

{\bf Hierarchy of anisotropic contributions.} In the presence of the
large-scale anisotropy (that is, the anisotropy introduced at scales of order
$L$ by the forcing), correlation functions of the model can be decomposed in
irreducible representations of the $d$-dimensional rotation group $SO(d)$.
Such a decomposition naturally arises from the corresponding OPE, provided
it is made in irreducible traceless tensor composite operators; the rank $l$
of a tensor operator can be used to label the terms of the $SO(d)$-expansion
and can be viewed as the measure of anisotropy of the corresponding term
(``sector''). Thus each anisotropic sector is characterized by its own
set of scaling exponents, the leading term is given by the $l$-th rank
composite operator with minimal critical dimension.

Explicit expressions for these dimensions were obtained to the first order
in $\eps$ and $\eta$. They reveal an hierarchy related to the degree of
anisotropy: the higher is the rank of the operator (the more anisotropic
is the contribution), the larger is the corresponding dimension, and thus
the less important is its contribution to the inertial-range behavior.

This hierarchy, expressed by the relations (\ref{hier2}) or (\ref{DER1}),
holds for all nontrivial scaling regimes of our model, all values of the
parameters $\alpha$, $\A$, $d$ and so on. It is similar to the hierarchy
relations derived earlier for the passive scalar \cite{RG3,RG4,ST} and
magnetic fields \cite{Lanotte2,Lanotte,AB} advected by the Gaussian
velocity ensembles.

In particular, this means that the overall leading term is given by the
exponent from the isotropic sector, ant it is therefore the same for the
isotropic and anisotropic forcing. It also should be stressed that the
independence of the scaling behavior in different sectors is a direct
consequence of the linearity of our model, independence of the exponents on
the random force, and the $SO(d)$ symmetry of the unforced model. On the
contrary, the {\it hierarchy} of the exponents follows from the explicit
expressions, obtained only by practical calculation.

According to the Kolmogorov--Obukhov theory \cite{Legacy,Monin}, the
anisotropy introduced at large scales by the forcing (boundary conditions,
geometry of an obstacle {\it etc}) dies out when the energy is transferred
down to smaller scales owing to the cascade mechanism (isotropization of
the developed turbulence in the inertial-range). The analytical
results discussed above confirm this classical concept and give a more
quantitative picture of the isotropization. The relevance of these results
for more realistic situations (scalar advected by the two-dimensional NS
field or the turbulent velocity itself) is briefly discussed below.

{\bf Effects of compressibility.} The anomalous exponents explicitly
depend on the parameter $\alpha\ge0$ that measures the compressibility
of the fluid. For the regimes determined by the fixed points $F^{+}$
(magnetic model with finite correlation time) and $R^{\A}$ (zero
correlation time, with additional inequalities for the parameter $\A$
satisfied by the magnetic case $\A=1$ and its vicinity), the hierarchy
of anisotropic contributions becomes less pronounced as $\alpha$ grows:
the anisotropic corrections in Eq. (\ref{struc2}) become closer to each
other and to the leading term as $\alpha$ grows. Thus the compressibility
enhances the penetration of the large-scale anisotropy into the inertial
range. The situation is opposite for all the other regimes, which arguably
can be attributed to the influence of the pressure term.

{\bf Effects of pressure.} The dependence on the parameter $\A$, that
controls the pressure effects, is essentially different from the
dependence on $\alpha$: for the regimes with nonzero correlation time,
the value of the corresponding invariant variable can take only discrete
values $\A_{*}=-1$, 1, $\alpha$. The behavior for the fixed point $\A_{*}=1$,
which corresponds to the magnetic case (``turbulence  without pressure''),
shows no serious difference from the regimes with pressure.
For the rapid-change limit, the exponents continuously depend on $\A$,
and the value of $\A=1$, where the pressure effects vanish, is not
distinguished either.

{\bf Relevance for the NS turbulence.}
The picture outlined above for passively advected fields (a superposition
of power laws with universal exponents and nonuniversal amplitudes) seems
rather general, being compatible with that established recently in the
field of NS turbulence, on the basis of numerical simulations of channel
flows and experiments in the atmospheric surface layer; see Refs.
\cite{Arad1,Arad2,Arad3,Arad4,BDAT,BCTT} and references therein.
It was shown that the leading terms of the inertial-range behavior are
the same for isotropic and anisotropic forcing \cite{Arad1,Arad2}.
In the papers \cite{Arad3,Arad4,BDAT,BCTT}, the velocity correlation
functions were decomposed in the irreducible representations
of the rotation group. It was argued that in each sector of the
decomposition, scaling behavior can be found with apparently universal
exponents. The amplitudes of the various contributions are nonuniversal,
through the dependence on the position in the flow, the local degree
of anisotropy and inhomogeneity, and so on.

This is rather surprising because the equations for the correlation
functions in such cases are neither closed nor isotropic and homogeneous.
Although the hierarchy similar to Eq. (\ref{hier2}) is demonstrated by
the critical dimensions of certain tensor operators in the stirred NS
turbulence, see Sec. 2.3 of \cite{turbo}, the
relationship between them and the anomalous exponents is not obvious there.
It is worth recalling here that the so-called ``additive fusion rules,''
hypothesized for the NS turbulence in a number of papers, Refs.
\cite{Falk1,GK,Eyink}, and characteristic of the models with multifractal
behavior (see Ref. \cite{DL}), arise naturally in the context of the models
of passive advection owing to their {\it linearity}. The existing results
for the Burgers turbulence can also be interpreted naturally as a consequence
of similar fusion rules, where only finite number of dangerous operators
contributes to each structure function, see Ref. \cite{Burg1}.

One can thus speculate that the anomalous scaling for the genuine turbulence
can also appear a linear phenomenon in the following sense. Let us split
the total velocity field into the two parts, the background field and the
perturbation (e.g., large-scale and small-scale, or soft and hard
components), linearize the original stochastic equation with respect to
the latter, choose an appropriate statistics for the former (e.g. Gaussian
distribution with Kolmogorov exponents, the description suggested for the
large-scale field by the experiment). Then the small-scale perturbation
field will show anomalous scaling behavior with nontrivial exponents,
which can be calculated systematically within a kind of $\eps$ expansion.
The corrections due to the nonlinearity can be treated perturbatively,
and if they appear irrelevant (e.g. in the sense of Wilson), they will
not affect the exponents calculated within the linearized model.
In such a case the passive vector field can give the anomalous exponents
for the NS velocity field exactly. In other words, such linearized model
will belong to the same universality class as the real NS equation,
like the simplified Ising or Heisenberg models are believed to belong
to the same universality class as real ferromagnets or binary alloys.
It thus might happen that the anomalous behavior of the real
inertial-range turbulence is exactly described by one of the
nontrivial fixed points for the passive vector model.

Of course, one should not insist too much on such a simple scenario for
the anomalous scaling, but it is worthy of attention. In this connection,
we could also recall that the passive vector field can indeed reveal the
anomalous exponents of the stochastic NS velocity field if the random
forcing of the former is chosen to be statistically correlated with that
of the latter; see \cite{BBT}.

{\bf Validity of the $\eps$ expansion and the applicability of the model.}
A serious question is that of the validity of the $\eps$ expansion
and the possibility of the extrapolation of the results, obtained
within the $\eps$ expansions, to the finite values $\eps=O(1)$.
For the rapid-change model, the $\eps$ expansion works surprisingly well.
It was shown \cite{RG100} that the knowledge of three terms allows one to
obtain reasonable predictions for finite $\eps\sim1$; even the plain $\eps$
expansion captures some subtle qualitative features of the anomalous
exponents established in analytical and numerical solutions of the exact
zero-mode equations and numerical experiments. The agreement can be
further improved by using special tricks (like the ``inverse'' $\eps$
expansion) or interpolation formulas \cite{RG100}.

In the case of the Gaussian model with a finite correlation time, however,
there is a natural upper bound for the range of validity of the results,
obtained within the $\eps$ expansion: for $\eps>1$ the velocity field
(and hence all its powers) become dangerous (its critical dimension
$\Delta_{v}=1-\eps$, known exactly due to the Gaussianity, becomes
negative). The spectrum of their dimensions is unbounded from below,
and in order to find the small-$mr$ behavior one has to sum up all their
contributions in the representations like (\ref{struc2}). This problem is
discussed in detail in \cite{RG3} for the passive scalar field; the infrared
perturbation theory was employed there to perform the required summation for
the pair correlation function, in the frozen regime, and within the one-loop
approximation for the Wilson coefficients. It was argued that, in that
special case, anomalous behavior is described by the same exponent below
and above the boundary $\eps=1$, but in general the problem remains open.

Physically, this is a manifestation of the fact that for $\eps>1$, the
so-called sweeping effects (kinematic transfer of the small-scale turbulent
eddies by the large-scale ones) become important. In a Galilean-covariant
problem such composite operators would not give any contribution into the
Galilean invariant quantities (structure functions), as it happens in the
RG approach to the stochastic NS equation; see the discussion in Refs.
\cite{JETP} and \cite{APS}. As was pointed out in Ref. \cite{OU}, the
Gaussian model with finite correlation time suffers from the lack of Galilean
invariance and therefore misrepresents the sweeping effects: they penetrate
into the correlation functions of the scalar and can lead to their strong
unphysical dependence on $L$. Therefore the value $\eps=1$ can also be viewed
as the threshold above which the model itself becomes unphysical. [To justify
the Gaussian model for $\eps>1$, however, one may recall that the results of
\cite{OU} show that it gives a reasonable description of the passive
advection in an appropriate frame, where the mean velocity field vanishes.]

We may therefore conclude that the next important step is the analytical
derivation of anomalous exponents of a passive scalar and vector quantities
advected by the Galilean covariant velocity ensemble, generated by the
stochastic NS equation; this work is now in progress.

\acknowledgments
The authors thank L.~Ts.~Adzhemyan, A.~Kupiainen, P.~Muratore-Ginanneschi,
A.~N.~Vasil'ev and D.~V.~Vassilevich for discussions.
The work of N.V.A. and J.H. was performed within the framework of the
Nordic Grant for Network Cooperation with the Baltic Countries and
Northwest Russia (Grant No.~FIN-6/2002).
M.H. and M.J. were supported by the Slovak Academy of Sciences
(Grant VEGA No.~3211).
N.V.A. was supported by the Academy of Finland
(Grant No.~79781) and the program ``Universities of Russia.''
N.V.A. thanks the Organizers of the Program on Developed Turbulence
(Vienna, 2002) and the Erwin Schr\"{o}dinger International Institute
for Mathematical Physics, where a part of this work was performed and
extensively discussed. Discussions with L.~Biferale, A.~Celani,
M.~Cencini, T.~Dombre, G.~Falkovich, K.~Gaw\c{e}dzki, A.~Mazzino,
P.~Olla, M.~Vergassola and D.~Vincenzi are especially acknowledged.

\begin{table}
\caption{Canonical dimensions of the fields and parameters in the
model (\protect\ref{action}).}
\label{table1}
\begin{tabular}{ccccccccc}
$F$ & $\theta$ & $\theta'$ & $ {\bf v} $ & $\nu$, $\nu _{0}$ & $m=1/L$,
$\mu$, $\Lambda$ & $g_{0}$ & $u_{0}$ &  $g$, $u$, $\A_0$, $\A$, $\alpha$ \\
\tableline
$d_{F}^{k}$ & 0 & $d$ & $-1$ & $-2$ & 1& $2\eps $ & $\eta$ & 0 \\
$d_{F}^{\omega}$ & $-1/2$ & $1/2$ & 1 & 1 & 0 & 0 & 0 & 0 \\
$d_{F}$ & $-1$ & $d+1$ & 1 & 0 & 1 & $2\eps $ & $\eta$ & 0 \\
\end{tabular}
\end{table}


\begin{figure}
\centerline{ \hbox{ \epsfig{file=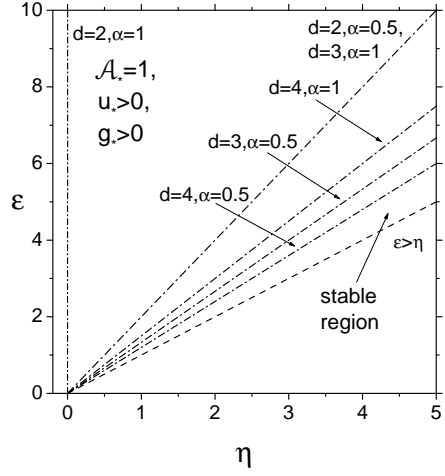,width=6cm}\hskip .5
cm}} \vspace{0.5cm} \caption{Regions of infrared stability of the
fixed point (\protect\ref{F+}) with a finite correlation time in
the asymptotic regime. Basins of attraction are shown for a
representative set of values of the space dimension $d$ and the
parameter $\alpha$ of the relative contribution of the
longitudinal part in the correlation function of the velocity
field. The region of stability for any indicated values of $d$ and
$\alpha$ lies between the dashed ray ($\varepsilon=\eta$, $\eta\ge
0$) and the correspondingly marked dash-dotted ray in the upper
half of the $\eta$, $\varepsilon$ plane. Note different scales on
the coordinate axes.} \label{Fig2}
\end{figure}

\begin{figure}
\centerline{ \hbox{ \epsfig{file=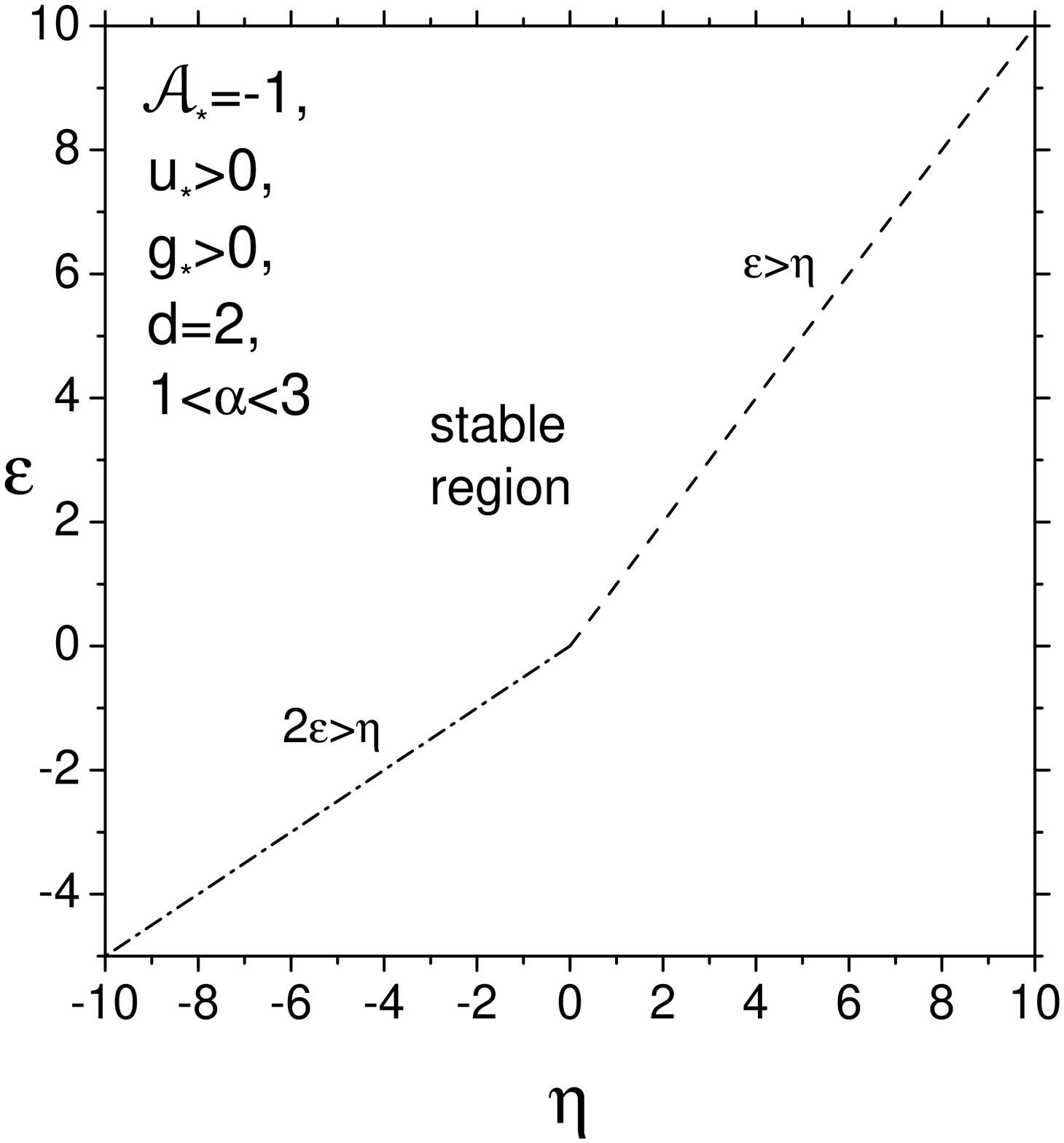,width=5cm}\hskip
.5 cm \epsfig{file=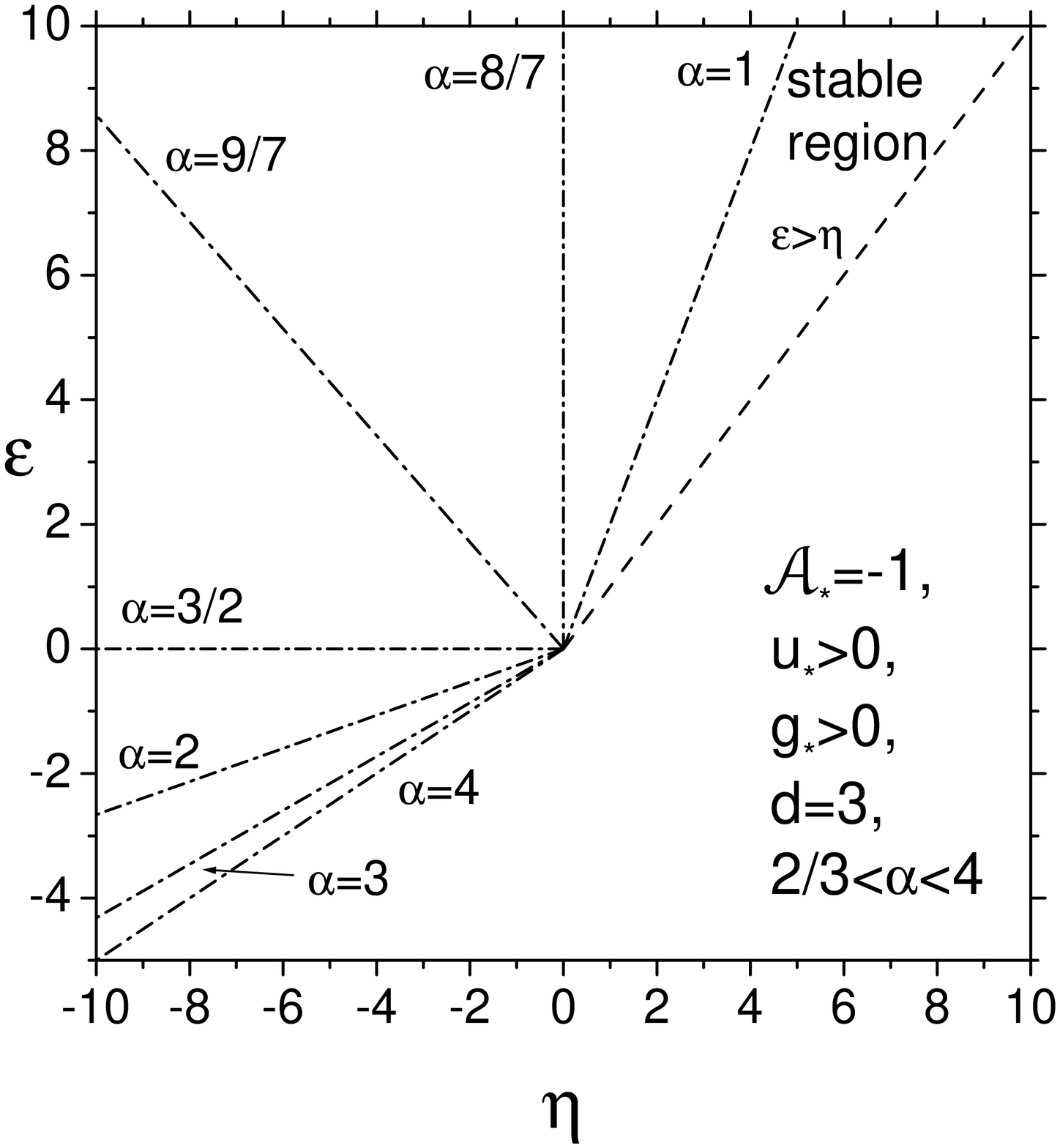,width=5cm}\hskip .5 cm
\epsfig{file=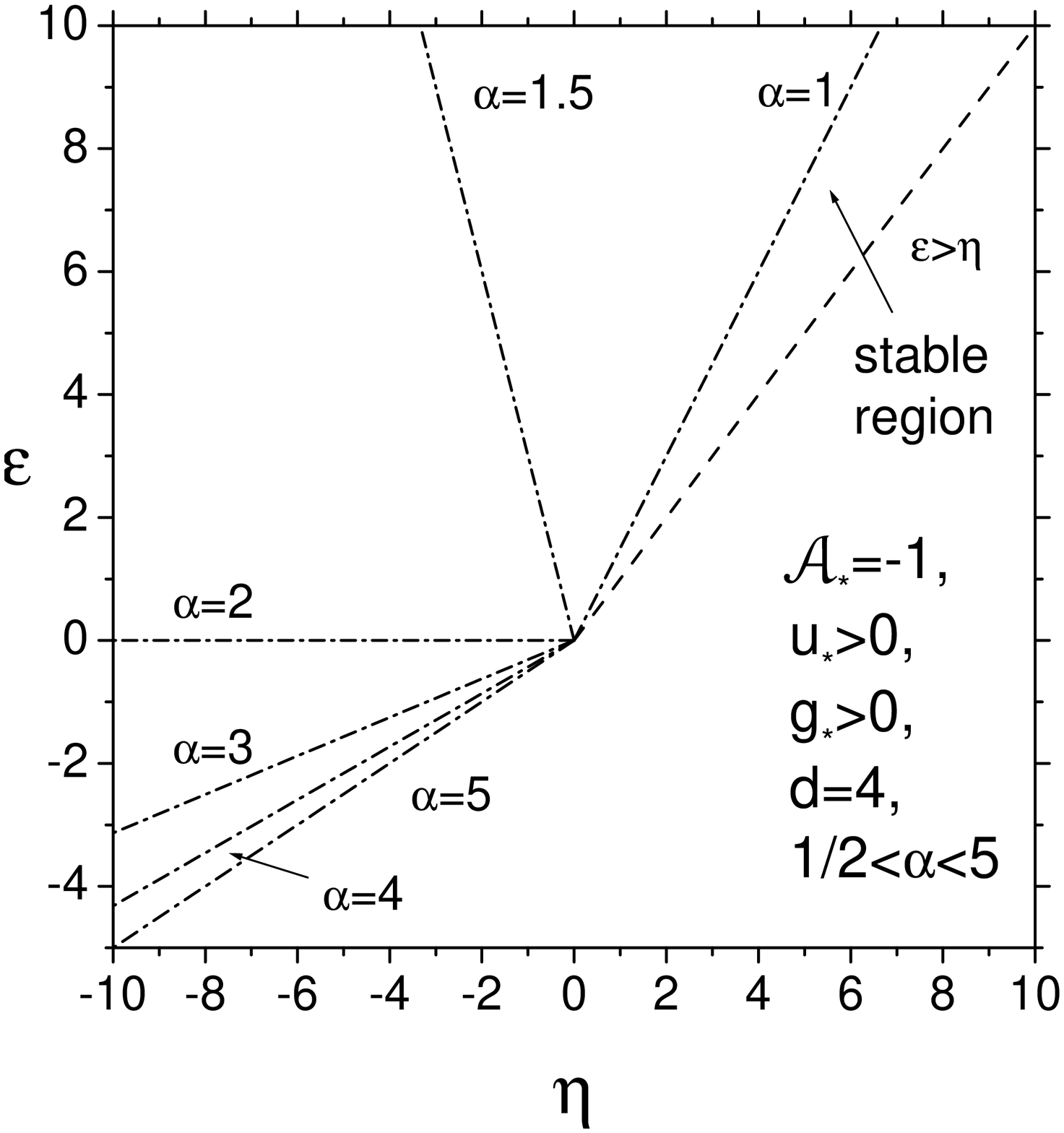,width=5cm}\hskip .5 cm}}
\vspace{0.5cm} \caption{Regions of infrared stability of the fixed
point (\protect\ref{F-}) with a finite correlation time in the
asymptotic regime. Basins of attraction of the three fixed points
are shown for a representative set of the longitudinal parameter
$\alpha$ in space dimensions two, three and four. The region of
stability for any indicated value of $\alpha$ lies from the dashed
ray ($\varepsilon=\eta$, $\eta\ge 0$) to the left up to the
correspondingly marked dash-dotted ray. Note different scales on
the coordinate axes. Contrary to the quenched-disorder case,
negative values of the correlation falloff parameter $\varepsilon$
are (formally) allowed.} \label{Fig3}
\end{figure}

\begin{figure}
\centerline{ \hbox{ \epsfig{file=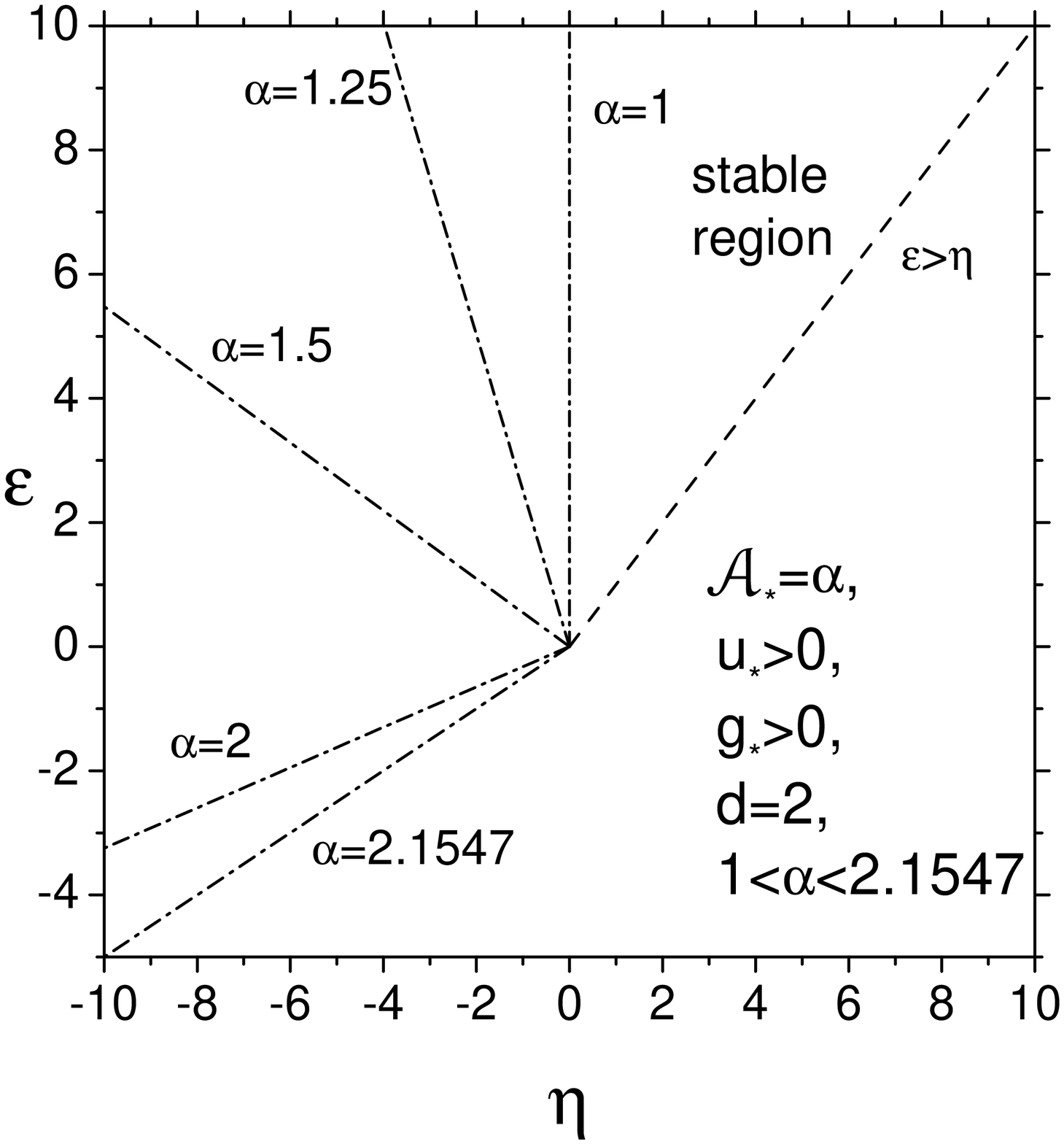,width=5cm}\hskip
.5 cm \epsfig{file=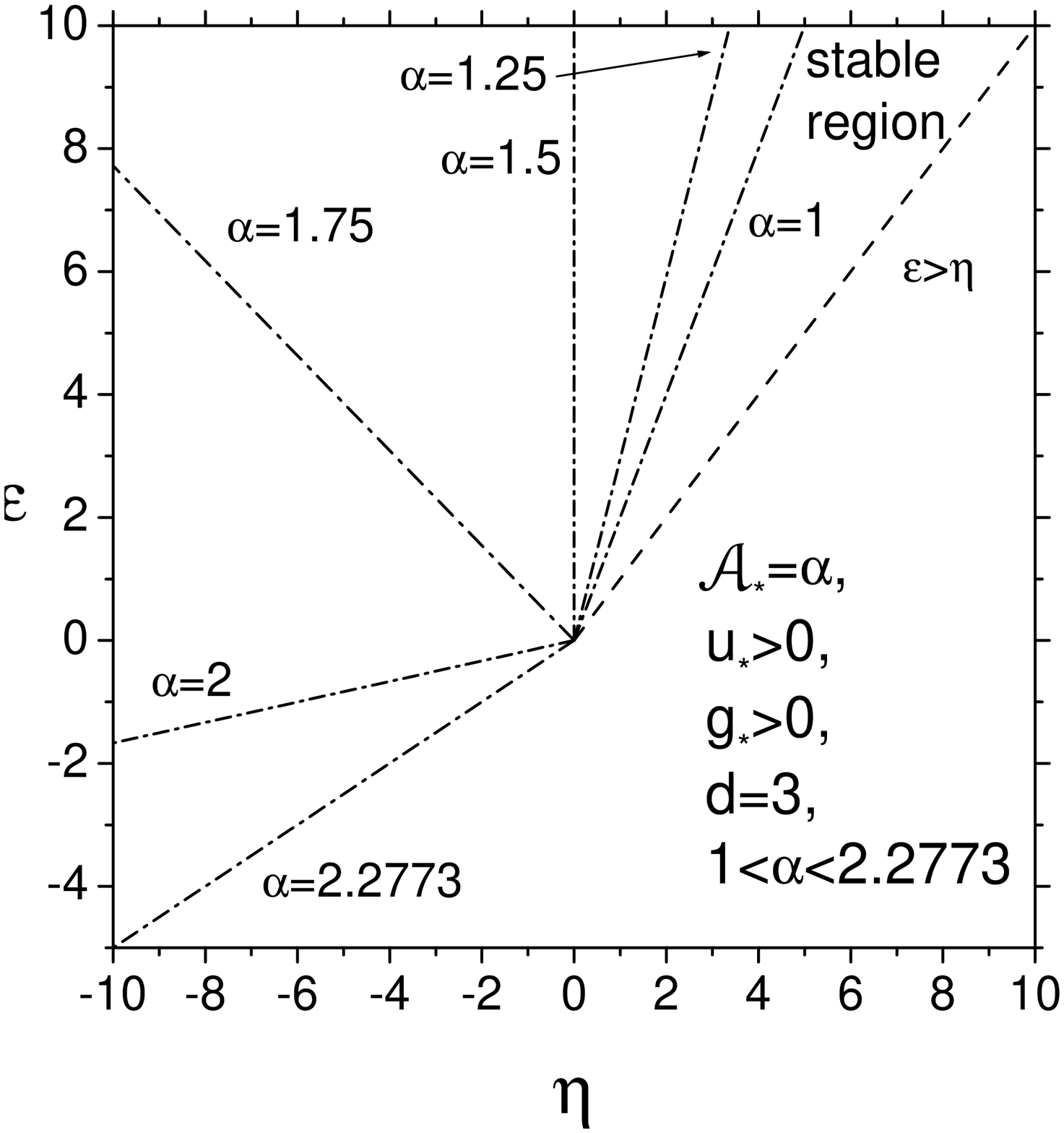,width=5cm}\hskip .5 cm
\epsfig{file=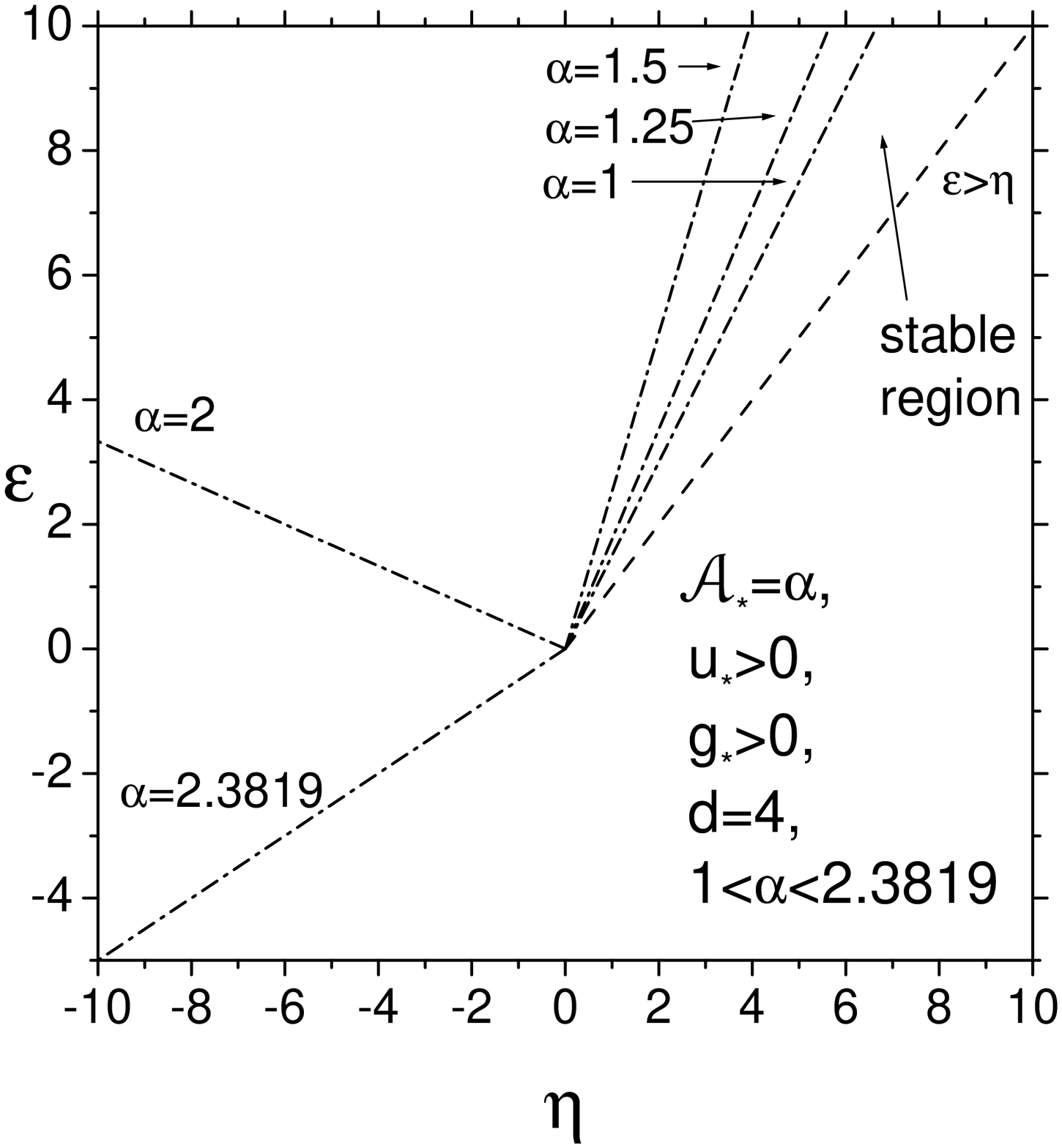,width=5cm}\hskip .5 cm}}
\vspace{0.5cm} \caption{Regions of infrared stability of the fixed
point (\protect\ref{Fa}) with a finite correlation time in the
asymptotic regime. Basins of attraction of the three fixed points
are shown for a representative set of the longitudinal parameter
$\alpha$ in space dimensions two, three and four. The region of
stability for any indicated value of $\alpha$ lies from the dashed
ray ($\varepsilon=\eta$, $\eta\ge 0$) to the left up to the
correspondingly marked dash-dotted ray. Note different scales on
the coordinate axes. Negative values of  $\varepsilon$ appear here
as well as for the fixed point (\protect\ref{Fa}).} \label{Fig4}
\end{figure}

\end{document}